\documentclass[12pt,amssymb,amsmath,tightenlines,showpacs]{article}
\usepackage[utf8]{inputenc}
\usepackage[a4paper,top=3cm,bottom=3cm,left=2cm,right=2cm]{geometry}
\usepackage{amssymb}
\usepackage{amsmath}
\usepackage{amsthm}
\usepackage{tipa}
\usepackage{latexsym} 
\usepackage{graphicx}
\usepackage{slashed}
\usepackage{bbold}
\usepackage{cancel}
\usepackage{color}
\usepackage{graphicx,epsfig,color}
\usepackage{cite}
\usepackage{tikz}
\usepackage{braket}
\usepackage[compat=1.1.0]{tikz-feynman}

\usepackage{caption}
\newcommand{\be}{\begin{equation}}
\newcommand{\ee}{\end{equation}}
\newcommand{\bea}{\begin{eqnarray}}
\newcommand{\eea}{\end{eqnarray}}

\numberwithin{equation}{section}

\begin{document}
	\graphicspath{{FIGURE/}}
	\topmargin=-1cm

\begin{flushright}
KA-TP-15-2026\\
P3H-26-048
\end{flushright}

\begin{center}

	{\Large{\bf 
Thermal Masses and Bubble-Wall Friction\\
\vspace{.5em}
in Cosmological Phase Transitions
}}\\

	\vspace*{0.8 cm}
	
{
Carlo Branchina$^a$, Stefania De Curtis$^b$, Luigi Delle Rose$^{c,d}$, Alessio Notari$^e$,\\ Giuliano Panico$^{f,b}$, and Matthew Starbuck$^g$
}
\vspace*{0.4cm}

{\footnotesize
{${}^{a}$\it Institute for Theoretical Physics, Karlsruhe Institute of Technology, \protect \\Wolfgang-Gaede-Str. 1,
76131 Karlsruhe, Germany}\\
\vskip 3pt
{${}^{b}$\it INFN Sezione di Firenze, Via G. Sansone 1, I-50019 Sesto Fiorentino, Italy}\\
\vskip 3pt
{${}^{c}$\it Dipartimento di Fisica, Università della Calabria, I-87036 Arcavacata di Rende, Cosenza, Italy}\\
\vskip 3pt
{${}^{d}$\it INFN, Gruppo Collegato di Cosenza, Arcavacata di Rende, I-87036, Cosenza, Italy}\\
\vskip 3pt
{${}^{e}$\it Dipartimento di Fisica, Sapienza University of Rome and INFN, Piazzale Aldo Moro 2, I-00185, Italy}\\
\vskip 3pt
{${}^{f}$\it Dipartimento di Fisica e Astronomia, \protect\\ Universit\`a di Firenze, Via G. Sansone 1, I-50019 Sesto Fiorentino, Italy}\\
\vskip 3pt
{${}^{g}$\it Department of Physics and Astronomy, University of Sussex, \protect\\ Brighton BN1 9QH, United Kingdom}\\
}

\vspace*{1 cm}

\begin{abstract}
\noindent
Bubble-wall friction controls the dynamics of first-order cosmological phase transitions. In Boltzmann-equation approaches, a major uncertainty arises from infrared gauge bosons, whose contribution is artificially enhanced in the massless approximation. We study the impact of thermal masses by including them consistently in both the Liouville operator and the collision integrals. Thermal masses suppress the source term for out-of-equilibrium perturbations while also reducing interaction rates. These effects largely cancel for top quarks, giving only percent-level changes, but they strongly suppress the infrared gauge-boson contribution, shifting the dominant momenta to scales of order the temperature. As a result, gauge bosons become subleading and wall velocities are close to those obtained from top-quark friction alone. We illustrate this in the singlet-extended Standard Model. Our results show that thermal masses reduce the sensitivity of friction calculations to the poorly controlled infrared sector of the plasma.
\end{abstract}
\end{center}

\vfill
\noindent\line(1,0){188}\\
{\scriptsize{E-mail: \texttt{carlo.branchina@kit.edu}, \texttt{stefania.decurtis@fi.infn.it}, \texttt{luigi.dellerose@unical.it}, \texttt{alessio.notari@uniroma1.it}, \texttt{giuliano.panico@unifi.it}, \texttt{m.starbuck@sussex.ac.uk}}
}

\thispagestyle{empty}

\newpage

\tableofcontents

\setcounter{equation}{0}
\setcounter{footnote}{0}
\setcounter{page}{1}

\section{Introduction}\label{introduction}

First-order phase transitions in the early Universe can give rise to a rich variety of observable signatures, including the generation of the matter-antimatter asymmetry, primordial magnetic fields, topological defects, dark-matter relics and, most notably, a stochastic background of gravitational waves. Future space-based interferometers are expected to be sensitive to gravitational waves produced during an electroweak-scale phase transition, opening a new observational window on the dynamics of electroweak symmetry breaking and on physics beyond the Standard Model.

The dynamics of a first-order phase transition is governed by the nucleation and expansion of bubbles of the stable phase within the surrounding metastable plasma. As the bubbles grow, the scalar fields forming the bubble wall interact with the plasma particles, driving them away from local thermal equilibrium. The resulting perturbations of the plasma back-react on the wall and generate a friction force that opposes its motion. The steady-state wall velocity is determined by the balance between this friction and the pressure difference that drives the phase transition. Since many phenomenological observables depend sensitively on the wall velocity, understanding the origin and magnitude of the friction is a central problem in the study of cosmological phase transitions.

The friction acting on the wall receives contributions both from the modification of the local-equilibrium properties of the plasma and from genuinely out-of-equilibrium effects. The latter are considerably more challenging to determine. The standard approach is to describe the relevant plasma species as quasiparticles whose distribution functions satisfy a Boltzmann equation in the background of the expanding wall. Solving this kinetic system provides the out-of-equilibrium perturbations, from which the friction can be computed and consistently incorporated into the equations governing the wall dynamics.

While this framework has been successfully applied in many studies~\cite{Moore:1995ua,Moore:1995si,Moore:2000wx,John:2000zq,Cline:2000nw,Fromme:2006wx,Megevand:2009gh,Huber:2013kj,Konstandin:2014zta,Kozaczuk:2015owa,Dorsch:2018pat,Cline:2020jre,Laurent:2020gpg,Friedlander:2020tnq,Wang:2020zlf,Dorsch:2021ubz,Cline:2021dkf,Lewicki:2021pgr,Dorsch:2021nje,DeCurtis:2022hlx,Laurent:2022jrs,DeCurtis:2022llw,DeCurtis:2022djw,Jiang:2022btc,DeCurtis:2023hil,Dorsch:2023tss,DeCurtis:2024hvh,Kainulainen:2024qpm,Branchina:2024rva,Ekstedt:2024fyq,Ai:2024btx,Dorsch:2024jjl,Branchina:2025adj,Barni:2025ifb,vandeVis:2025plm,Ekstedt:2025awx,Krajewski:2026kcm,Laurent:2026jvx}, its validity is expected to become less robust in the infrared region of the plasma. This issue is particularly relevant for gauge bosons. In the commonly adopted massless approximation, the Bose-Einstein distribution strongly enhances the contribution of soft $W$ modes $|\vec p\,| \lesssim g_W T$, which can provide a sizeable contribution to the total out-of-equilibrium friction. At the same time, these are precisely the modes for which the quasiparticle picture underlying the Boltzmann description is expected to be least reliable.

A natural question is therefore whether the large gauge-boson contribution found in previous analyses is a genuine physical effect or is partly driven by the treatment of the plasma as made of massless species.

Previous studies~\cite{Moore:2000wx,Arnold:1996dy,Huet:1996sh,Son:1997qj,Bodeker:1998hm,Bodeker:1999ud,Bodeker:1999ey,Arnold:1999jf,Arnold:1999ux} tried to address this issue by exploiting an alternative description of the soft gauge boson modes, which are described through a Langevin-type equation taking into account the screening effects of the plasma. This description allows for an extension of the range of validity of the effective theory down to $|\vec p\,| \sim g_W^2 T$. However, the separation between the scales $g_W T$ and $g_W^2 T$ is not parametrically large in the electroweak plasma, where $g_W$ is an ${\mathcal O}(1)$ coupling, and the gain is thus modest. It is therefore not obvious that the soft modes providing the dominant contribution to the friction can be cleanly isolated from the rest of the plasma dynamics.

In this work we revisit the problem from a different perspective. Rather than introducing a separate effective description for the infrared sector, we retain the Boltzmann treatment of the plasma dynamics and incorporate screening effects through thermal masses in the local-equilibrium distributions. Thermal masses modify both the source term generated by the Liouville operator and the collision integrals governing thermalisation. Their impact has usually been neglected in studies of bubble-wall friction, with the exception of Ref.~\cite{vandeVis:2025plm}, where thermal masses were included only in the Liouville operator.

We apply this framework to the singlet-extended Standard Model and determine the impact of thermal masses on the wall velocity and friction profiles. We find that thermal masses have only a modest effect on the top-quark contribution, owing to a compensation between their impact on the source and collision terms. In contrast, they strongly suppress the infrared enhancement of the $W$-boson contribution, rendering the gauge-boson friction subleading throughout the parameter space explored. As a result, friction calculations become significantly less sensitive to the poorly controlled infrared sector of the plasma, providing a more robust determination of the bubble-wall velocity.

The paper is organised as follows. In Sec.~\ref{sec:Boltzmann} we briefly review the Boltzmann description of plasma perturbations and outline the numerical strategy used to solve the resulting equations, based on Ref.~\cite{DeCurtis:2022hlx}. In particular, we summarise the spectral-decomposition method developed in Ref.~\cite{DeCurtis:2023hil,DeCurtis:2024hvh} for the treatment of the collision integrals. In Sec.~\ref{sec:kernels} we analyse how thermal masses modify the collision kernels. We then apply this framework to the singlet-extended Standard Model in Sec.~\ref{sec:application}, where we first discuss in detail two benchmark points, showing how the inclusion of thermal masses regulates the infrared sector of the plasma, and then perform a broader exploration of the model parameter space. Finally, Sec.~\ref{sec:conclusions} contains our conclusions.

\section{Boltzmann equation in a bubble-wall background}\label{sec:Boltzmann}

\subsection{Setting up the Boltzmann equation}

The starting point of our analysis is the integro-differential Boltzmann equation, which describes the evolution of the distribution functions of the particle species in the plasma, in the presence of a planar bubble wall. Setting the wall to be orthogonal to the $z$-direction, the distribution function $f_i$ of the $i$-th particle species satisfies the equation
\begin{equation}
\mathcal L[f_i]\equiv \left(\frac{p_z}{E_p}\partial_z-\frac{\left(m_i^2(z)\right)'}{2 E_p}\partial_{p_z}\right) f_i = - \mathcal C_i[f]\,.
\end{equation}
In this equation $\mathcal C_i[f]$ denotes the collisional operator, which describes the (local) interactions among plasma particles, whose main effect is to drive the system towards local thermal equilibrium (LTE).

To study the out-of-equilibrium corrections, we expand the distribution functions around local equilibrium, $f_i=f_{0,i}+\delta f_i$, with 
\begin{equation}
f_0=\frac{1}{e^{\beta p^\mu u_\mu}\pm 1}\,,
\end{equation}
where $u_\mu$ is the local four-velocity of the fluid.
The linearised Boltzmann equation reads
\begin{equation}
	\label{eq: linearised Boltzmann}
\mathcal L\left[\delta f_i\right] = \frac{p_z}{E_p}\mathcal S_i - \overline{\mathcal C}_i[\delta f]\,,
\end{equation}
where $\overline{\mathcal C}_i$ indicates the linearised collision integral, while $\mathcal S_i$ corresponds to a source term given by the action of the Liouville operator $\mathcal L$ on the local equilibrium distribution function $f_{0,i}\,$:
\begin{equation}
\frac{p_z}{E_p} \mathcal S_i\equiv \mathcal L\left[f_{0,i}\right]\,.
\end{equation}

Since the Liouville operator contains only first-order derivatives, the Boltzmann equation can be solved by applying the method of characteristics as proposed in Ref.~\cite{DeCurtis:2022hlx}. The characteristic curves are easily found by solving the equations
\begin{align}
\frac{dz}{ds}=1\,, 
\qquad \qquad 
\frac{dp_z}{ds}=-\frac{(m^2_i)'}{2p_z}\,,
\qquad \qquad 
\frac{dp_\perp}{ds}=0\,,
\end{align}
so that in the collisionless limit particles follow trajectories which preserve the transverse momentum $p_{\perp}$ and the longitudinal energy combination $p_z^2+m_i^2(z)$.
These define three kinds of flow paths: (i) particles moving toward the wall that have enough momentum $p_z$ to enter inside the bubble, (ii) particles moving toward the wall that do not have enough momentum, and are thus reflected from the wall, and (iii) particles emerging from the bubble. Details of the flow paths of the Liouville operator $\mathcal L$ were given in Ref.~\cite{DeCurtis:2022hlx} for a $z$-dependent mass of the form $m(z) = \lambda\, \phi(z)$, with $\lambda$ a generic coupling and $\phi(z)$ the field acquiring a non-trivial vacuum expectation value (VEV) $\braket{\phi}=\phi_0$ in the transition, parametrised as $\phi(z)=\phi_0(1+\tanh z/L)/2$. The extension to a more general mass term
including the coupling to several scalar fields and a thermal mass is straightforward. 

Since the collision operator $\overline{\mathcal C}_i[\delta f]$ introduces a dependence on the integral of $\delta f$,
a full solution of the linearised Boltzmann equation cannot be found in closed form. 
However, as shown in Ref.~\cite{DeCurtis:2022hlx}, a solution can be obtained through an iterative approach by successive approximations. This is achieved by estimating the integral contributions of the collisional terms using the solution obtained at the previous iterative step, thus obtaining (at each step) a regular differential equation that can be solved through the method of characteristics. This approach proves efficient and typically converges after only a few iterations.

In our analysis we will include in the collisional terms only $2 \to 2$ processes, which are expected to provide the dominant contributions to the thermalisation process.
Within this approximation the collision integral reads
\begin{equation}
\label{eq: collision integral}
\mathcal C_i[f]=\sum_j\frac{1}{4 N_p E_p} \int \frac{d^3 \vec k\, d^3\vec {p'}\,d^3\vec{k'}}{(2\pi)^5 2E_k \,2 E_{p'}\, 2 E_{k'}}\left|\mathcal M_{i,j}\right|^2 \delta^4(p+k-p'-k')\mathcal P_i[f]\,, 
\end{equation}
where $N_p$ and $E_p$ are the number of degrees of freedom and the energy of the given particle $i$, and the sum extends over all processes involving the particle in the initial or final state ($\mathcal M_{i,j}$ denotes the corresponding amplitudes).
The population factor $\mathcal P_{i}[f]$ is given by
(the $\pm$ signs correspond to bosons and fermions respectively)
\begin{equation}
\mathcal P_i[f]= f_i(p)f(k)\left(1\pm f(p')\right) \left(1\pm f(k')\right)-f(p')f(k')\left(1\pm f_i(p)\right)\left(1\pm f(k)\right)\,, 
\end{equation}
where distribution functions without a subscript are those of the particles appearing in the specific $\mathcal M_{i,j}$ process.

At linear order in $\delta f$, $\overline {\mathcal P}_i$ becomes 
\begin{equation}
\label{eq: linearised population}
\overline{\mathcal P}_i=f_{0,i}(p)f_0(k)\left(1\pm f_0(p')\right)\left(1\pm f_0(k')\right)\sum_{l\in \left(p,\,k,\,p',\,k'\right)}\frac{\mp \delta f(l)}{\left(f_0(l)\right)'}\,.
\end{equation}
The $\mp$ signs in the last factor correspond to incoming particles ($l=p,k$) and outgoing particles ($l=p',k'$), respectively.

The linearised collision integral can be split in two terms with different structures:
\begin{equation}\label{eq:coll_lin}
\overline{\mathcal C}_i = -\frac{f_{0,i}(p)}{f'_{0,i}(p)}\frac{\mathcal Q_i}{E_p} \delta f_i(p)  -f_{0,i}(p)\left(\frac{\braket{\delta f}_{i,a} - \braket{\delta f}_{i,s}}{E_p}\right)\,.
\end{equation}
The first term contains $\delta f_i(p)$ as a multiplicative factor and
\begin{equation}
   \mathcal Q_i\equiv  \frac{1}{4N_p}\frac{1}{(2\pi)^5}\int d^3 \vec k\, d^3\vec {p'}\,d^3\vec{k'} \sum_j \overline{\mathcal R}_{ij}\,,
\end{equation}
with
\begin{equation}
\overline{\mathcal R}_{ij} \equiv   \left|\mathcal M_{i,j}\right|^2  \left[\frac{f_0(k) (1\pm f_0(p'))(1\pm f_0(k'))}{2E_k 2E_{p'}2E_{k'}}\right]_j\,\delta^4(p+k-p'-k')\,.
\end{equation}
The subscript outside the square brackets has been added to stress that the equilibrium distribution functions and the energy factors inside of it depend on the specific process $j$.
On the other hand, in the second term in eq.~(\ref{eq:coll_lin}), $\delta f_i$ appears under the integral sign:
\begin{align}
\braket{\delta f}_{i,a} &\equiv  \frac{1}{4N_p (2\pi)^5} \int d^3 \vec k\, d^3\vec {p'}\,d^3\vec{k'} \sum_j \overline{\mathcal R}_{ij} \left(\frac{\delta f(k)}{f'_0(k)}\right)_j\,, \\
\braket{\delta f}_{i,s} &\equiv  \frac{1}{4N_p (2\pi)^5}\int d^3 \vec k\, d^3\vec {p'}\,d^3\vec{k'}\sum_{l=\{p',k'\}}\sum_{j} \overline{\mathcal R}_{ij} \left(\frac{\delta f(l)}{f'_0(l)}\right)_j\,.
\end{align}
Due to their different kinematics, we separate the contributions arising from annihilation processes, $\braket{\delta f}_{i,a}$, and scattering processes, $\braket{\delta f}_{i,s}$.
Defining the full bracket $\braket{\delta f}$ as the difference between the annihilation and scattering one, $\braket {\delta f}_i \equiv \braket {\delta f}_{i,a} - \braket {\delta f}_{i,s}$, the Boltzmann equation can be written in a compact form as 
\begin{equation}
\mathcal L\left[\delta f_i\right] - \frac{f_{0,i}(p)}{f'_{0,i}(p)}\frac{\mathcal Q_i}{E_p} \delta f_i = \frac{p_z}{E_p}\mathcal S_i + \frac{f_{0,i}(p)}{E_p}\braket{\delta f}_i\,.  
\end{equation}
Along the flow paths, this reduces to 
\begin{equation}\label{eq:Boltzmann_paths}
\left(\frac{d}{dz} - \frac{f_{0,i}}{f'_{0,i}}\frac{\mathcal Q_i}{p_z}\right) \delta f_i = \mathcal S_i + \frac{f_{0,i}}{p_z}\braket{\delta f}_i\,.  
\end{equation}

Independently of the specific approach chosen to solve the Boltzmann equation, the evaluation of the collision integral represents the challenging part of the whole calculation. Within the iterative procedure, the quantity $\mathcal{Q}$ does not depend on $\delta f$ and can therefore be computed once and for all. The brackets, on the contrary, must be evaluated at every iteration step and thus constitute the computational bottleneck.
A method based on a spectral decomposition of the collision kernel, which we will briefly review in the next section, was developed in Ref.~\cite{DeCurtis:2023hil}. It reduces the task to the combination of a single a-priori computation and a one-dimensional integration to be performed at each step. 

As commonly done in the literature, to reduce the computational complexity, we adopt a set of approximations. As we already mentioned, we only consider $2 \to 2$ processes.
We simplify the amplitudes including only the leading-log approximation, which takes into account only the $t$ and $u$ channel contributions.
Moreover, in the matrix elements entering the collision integrals, all external particles are treated as massless, while thermal masses are retained only in the propagators of the exchanged states to regulate infrared divergences.
The amplitudes for the relevant processes involving the top quark and the electroweak gauge bosons are reported in Table~\ref{tab:amplitudes}. The masses that we include in the propagators are the thermal masses of the corresponding species~\cite{Moore:1995si,Weldon:1982bn}, electric masses for the gauge bosons and fermionic thermal masses, as follows:
\begin{equation}
 m_q^2=\frac{g_s^2}{6} T_n^2\,, \qquad  m_{g, \textsc{D}}^2=2g_s^2 T_n^2\,, \qquad m_{W, \textsc{D}}^2= \frac{5 g_{_W}^2}{3}T_n^2\,, \qquad m_l^2= \frac{3g_{_W}^2}{32} T_n^2\,. 
\end{equation}

\begin{table}[t]
    \centering
    {\small
    \begin{tabular}{c|c}
        process & $|{\cal M}|^2$\\
        \hline
        \rule{0pt}{1.75em}$t \bar t \to gg$ & $\displaystyle \frac{128}{3} g_s^4 \left[ \frac{ut}{(t - m_q^2)^2} +  \frac{ut}{(u- m_q^2)^2} \right ]$\\
        \rule{0pt}{1.75em}$tg \to tg$ & $\displaystyle- \frac{128}{3} g_s^4 \frac{su}{(u-m_q^2)^2} + 96 g_s^4 \frac{s^2 + u^2}{(t - m_{g, \textsc{d}}^2)^2}$\\
        \rule{0pt}{1.75em}$tq \to tq$ & $\displaystyle160 g_s^4 \frac{s^2 + u^2}{(t - m_{g, \textsc{d}}^2)^2}$\\
        \hline
        \rule{0pt}{1.75em}
        $Wq \to qg$ & $-144g_s^2 g_{_W}^2 \displaystyle \frac{s t}{(t-m_q^2)^2}$\\
        \rule{0pt}{1.75em}
        $Wg \to \bar q q$ &
        $144g_s^2 g_{_W}^2 \displaystyle \frac{u t}{(t-m_q^2)^2}$\\
        \rule{0pt}{1.75em}
        $WW \to \bar f f$ & $\displaystyle 27g_{_W}^4 \left[ \displaystyle\frac{3 ut }{(t-m_q^2)^2} \delta_{fq}+ \displaystyle\frac{ut }{(t-m_l^2)^2}\delta_{fl}\right]$\\
        \rule{0pt}{1.75em}
        $W f \to W f$ & $\left(3\,\delta_{fq}+\delta_{fl}\right) 72 g_{_W}^4 \displaystyle\frac{s^2+u^2}{(t-m_{\textsc{w}, {\textsc{d}}}^2)^2} - \displaystyle 27 g_{_W}^4 \left[\displaystyle\frac{3 st}{(t-m_q^2)^2} \delta_{fq} + \displaystyle\frac{st}{(t-m_l^2)^2}\delta_{fl}\right]$\\
    \end{tabular}
    }
    \caption{Amplitudes for the processes relevant to the top quark (first three rows) and the $W$ bosons (last four rows) in the leading-log approximation. In the process $t q \to t q$, the sum over all light quarks and antiquarks has been performed. The subscript $l$ denotes leptons, while $q$ denotes light quarks in the top-quark amplitudes and all quarks in the $W$-boson amplitudes. In the first column, $f$ generically denotes either a quark or a lepton.}\label{tab:amplitudes}
\end{table}

\subsection{Collision integrals and spectral decomposition}

As we already mentioned, the computation of the bracket contributions $\braket{\delta f}$ is a critical step in the solution of the Boltzmann equation. In this section we discuss how this task can be performed efficiently through a suitable reformulation of the collision integrals~\cite{DeCurtis:2023hil}.

To illustrate the procedure, we focus on the top-quark contributions. In this case the dominant processes contributing to thermalisation are annihilation into a gluon pair $t\bar t\to gg$, and scattering onto gluons $tg\to tg$ and light quarks $tq\to tq$.

The first step in the computation of the bracket is to rewrite it as the action of an integral kernel $\mathcal K_{pq}$ on the density perturbation $\delta f_t(q)$:
\begin{equation}
\braket{\delta f}_t = \frac{1}{4 N_p} \int \frac{d^3 \vec q}{2E_q}  f_{0,t}(q) \,\mathcal K_{pq}\,\frac{\delta f_t(q)}{\left(f_{0,t}(q)\right)'}\,.
\label{eq: bracket kernel}
\end{equation}
The kernel depends on both the external momentum $p$ appearing in the Boltzmann equation and the integration momentum $q$, therefore it depends on $|\vec p\,|$, $|\vec q\,|$ and the angle between the two vectors $\cos \theta_{pq}$: $\mathcal K_{pq}\equiv \mathcal K(|\vec p\,|,|\vec q\,|,\cos\theta_{pq})$. Separating the contributions from annihilation and scattering, we have $\mathcal K_{pq} \equiv \mathcal K_{pq}^a-\mathcal K_{pq}^s$
(one can easily recognise $\braket{\delta f}_a$ and $\braket{\delta f}_s$ here)   
\begin{align}
 \mathcal K_{pk}^a &=\frac{1}{(2\pi)^5}\int d^3\vec{p'} \,d^3\vec k' \sum_{j\in \{j_a\}}|\mathcal M_{t,j}|^2\, \frac{(1\pm f_0(p'))(1\pm f_0(k'))}{2E_{p'}\,2E_{k'}}\Bigg|_j \, \, \delta^4(p+k-p'-k')\,,\label{eq:Ka_top}\\
 \mathcal K_{pp'}^s &=\frac{1}{(2\pi)^5}\int d^3\vec k \,d^3\vec k'\sum_{j\in \{j_s\}}|\mathcal M_{t,j}|^2\, \frac{e^{\beta p'^\mu u_\mu}f_0(k)(1\pm f_0(k'))}{2E_k\,2E_{k'}}\Bigg|_j\,\, \delta^4(p+k-p'-k')\,,\label{eq:Ks_top}
\end{align}
with the indices $j_a$ and $j_s$ running over processes with a second top quark state in the initial state ($t \bar t\to gg$ in this case) and processes with a top quark in the final state ($tg\to tg$ and $tq\to tq$), respectively.

The angular dependence in $\mathcal K_{pq}$ can be factored out using a Legendre decomposition
\begin{equation}
    \mathcal K_{pq}(|\vec p\,|,|\vec q\,|,\cos\theta_{pq})= \sum_{l=0}^\infty \frac{2l+1}{2} \mathcal G_l(|\vec p\,|,|\vec q\,|)P_l(\cos\theta_{pq})\,.
\end{equation}
Similarly, writing $\vec p$ in spherical coordinates, we decompose the perturbation 
\begin{equation}
    \delta f_t(\vec p, z)= \sum_{l=0}^\infty \frac{2l+1}{2}\psi_{l,t}(|\vec p\,|,z) P_l(\cos\theta_p)\,,
\end{equation}
with $\cos\theta_p$ being related to $\cos\theta_{pq}$ as $\cos\theta_{pq}=\cos\theta_p\cos\theta_q+\sin\theta_p\sin\theta_q\cos(\phi_q-\phi_p)$. This relation can be used to perform the integration over $\phi_q$ in eq.~\eqref{eq: bracket kernel}. In terms of the Legendre components, the bracket becomes
\begin{equation}
\braket{\delta f}_t = \frac{\pi}{4 N_p}\int \mathcal D q\,\sum_{l=0}^\infty \frac{2l+1}{2}\mathcal G_{l,pq}\,\frac{\psi_{l,t}(q)}{\left(f_{0,t}(q)\right)'}P_l(\cos\theta_p)\,,
\end{equation}
with $\mathcal Dq \equiv f_{0,t}(q) |\vec q\,|^2dq/E_q$ and $\mathcal G_{l,pq}\equiv \mathcal G_l(|\vec p\,|,|\vec q\,|)$.

With the above definitions, the kernels $\mathcal K^a_{pq}$ and $\mathcal K^s_{pq}$ (and in turn their Legendre components $\mathcal G_{l,pq}$) are symmetric under the exchange $p\leftrightarrow q$. This is obvious for the annihilation processes, since the symmetry simply exchanges the two incoming particles ($p\leftrightarrow k$ in eq.~(\ref{eq:Ka_top})). For the scattering processes, instead, the incoming and outgoing tops are exchanged, namely $p \leftrightarrow p'$ in eq.~(\ref{eq:Ks_top}). The invariance can be seen exchanging $k \leftrightarrow k'$ and noticing that
\begin{equation}\label{eq:symm_momentum}
 e^{\beta p'^\mu u_\mu} f_0(k) (1 \pm f_0(k')) = e^{\beta p'^\mu u_\mu} e^{\beta k'^\mu u_\mu} f_0(k) f_0(k') = e^{\beta p^\mu u_\mu} e^{\beta k^\mu u_\mu} f_0(k) f_0(k')\,,
\end{equation}
thanks to four-momentum conservation.

The symmetry of the kernel allows us to interpret $\braket{\delta f}_t$ as the action of a Hermitian operator $\mathcal O[g]=\sum_l \mathcal O_l[g]$ on the perturbation $\delta f_t$,\footnote{The corresponding scalar product is
\begin{equation}\label{eq:sc_product}
f\cdot g = \int \mathcal Dq \,f(|\vec q\,|)\, g(|\vec q\,|)\,.
\end{equation}
} with the Legendre block operators
\begin{equation}
    \mathcal O_l[g] = \int \mathcal D q\,\, \mathcal G_{l,pq}\,  g(|\vec q\,|) 
\end{equation}
being Hermitian themselves. Each operator admits a spectral decomposition into its natural basis of eigenfunctions $\{\zeta_{l,i}\}$ in terms of the eigenvalues $\{\lambda_{l,i}\}$,
\begin{equation}
\mathcal G_{l,pq}=\sum_i \lambda_{l,i}\, \zeta_{l,i}(|\vec p\,|)\zeta_{l,i}(|\vec q\,|).
\end{equation}
One finally gets 
\begin{equation}
\braket{\delta f}_t = \frac{\pi}{4 N_p}\sum_{l=0}^\infty  \sum_i \frac{2l+1}{2} \lambda_{l,i}\, \zeta_{l,i}(|\vec p\,|) P_l(\cos\theta_p)\int \mathcal D q\,\zeta_{l,i}(|\vec q\,|)\frac{\psi_{l,t}(q)}{\left(f_{0,t}(q)\right)'}\,. 
\label{eq: bracket decomposed}
\end{equation}

The procedure extends straightforwardly to additional particle species, since the diagonalisation of each element $\mathcal K$ proceeds analogously.

In our analysis, besides the top quark, we also include the out-of-equilibrium effects of the $W$ bosons\footnote{For simplicity, we neglect the contributions associated with the hypercharge symmetry $U(1)_Y$ and therefore treat the $W$ and $Z$ bosons as a degenerate triplet.}. In the presence of multiple out-of-equilibrium species, cross terms generally appear in the bracket contributions. This occurs in particular for the top-quark and $W$-boson sectors. Although such terms could in principle be incorporated into the analysis, doing so would couple the Boltzmann equations for the top quarks and $W$ bosons.
To avoid this additional complexity, we neglect the cross terms and instead describe the top quarks and $W$ bosons through two independent Boltzmann equations. This approximation is justified a posteriori, since the out-of-equilibrium contribution of the $W$ bosons is found to be significantly smaller than that of the top quarks once thermal masses are included.

\subsection{Inclusion of thermal masses}

Since departures from local thermal equilibrium are small, thermal effects can be incorporated by replacing the free-particle dispersion relations with their finite-temperature counterparts. This affects all ingredients entering the Boltzmann equation. The equilibrium distributions are modified, changing the source term $\mathcal S_i$. The effective masses entering the Liouville operator are altered, and the collision operator $\mathcal C_i$ receives corrections through both the scattering kinematics and the interaction rates.

The finite-temperature dispersion relations of scalar and fermionic fields are relatively simple (see eg.~\cite{Kraemmer:1994az,Bellac:2011kqa}). Even for massless fields, thermal effects generate an effective mass, producing a finite energy gap for infrared modes.\footnote{In our analysis we neglect the possible contributions from additional fermionic quasiparticles (plasminos).} The resulting thermal mass exhibits a mild momentum dependence, being slightly reduced in the infrared region.

Gauge bosons display a richer structure (see eg.~\cite{Bellac:2011kqa}), as transverse and longitudinal polarizations obey different dispersion relations. The transverse modes acquire a thermal mass that is slightly smaller at low momentum and approaches its asymptotic value in the ultraviolet. The longitudinal modes, corresponding to plasmons, are also massive but contribute appreciably only for momenta $|p| \lesssim T$. Their effective mass coincides with that of the transverse modes at vanishing momentum and decreases as the momentum increases.

To keep the numerical analysis tractable, we approximate the finite-temperature dispersion relations by constant effective masses. We use the asymptotic thermal masses evaluated in the symmetric phase at the nucleation temperature $T_n$, namely
\begin{equation}\label{eq:th_masses}
 m_q^2=\frac{g_s^2}{6} T_n^2 \qquad  m_g^2=g_s^2\, T_n^2, \qquad m_{_W}^2= \frac{11}{12}g_{_W}^2 T_n^2, \qquad m_l^2= \frac{3g_{_W}^2}{32} T_n^2\,. 
\end{equation}
For the $W$ bosons, transverse and longitudinal polarizations are treated identically.
As will be discussed below, this approximation is justified by the fact that the dominant out-of-equilibrium contributions to the friction originate from particles with momenta $|p|\sim T$. In this momentum range, the thermal masses vary only moderately and can be regarded as approximately constant.

The approximation of constant thermal masses leads to an important simplification: the force term appearing in the Liouville operator is unchanged
\begin{equation}
(m_i^2(z))' \rightarrow \bigl(m_i^2(z)+m_{\rm th}^2\bigr)'
= (m_i^2(z))'\,.
\end{equation}
As a consequence, the characteristic trajectories in phase space are not modified.

Moreover, the collision operator retains essentially the same structure as in the absence of thermal masses. In particular, if one further adopts the approximation of massless kinematics in the matrix elements and phase-space integrals, the bracket operator remains independent of $z$. Its spectral decomposition can therefore be performed once and subsequently reused throughout the computation.

A minor modification is nevertheless required. When thermal masses are included in the equilibrium distribution functions while massless kinematics is adopted in the collision integrals, the relations that guarantee the symmetry of the scattering kernels are no longer satisfied exactly. In particular, Eq.~(\ref{eq:symm_momentum}) relies on the same energy-momentum relation being used both in the equilibrium distributions and in the conservation laws entering the collision operator. Once thermal masses are included only in the former, the identity is violated. To restore the symmetry, we adopt a simple interpolation for the energy entering the equilibrium distributions,
\begin{equation}
p^0=\sqrt{|\vec p\,|^2+m^2}
\simeq |\vec p\,|+a\,m\,.
\end{equation}
This prescription is just a technical device that preserves the Hermiticity of the collision operator while retaining the computational advantages of the massless-kinematics approximation. Since the resulting friction is sensitive to the precise value of  the parameter $a$ only at the sub-percent level, we simply fix $a=1/2$ for all fields. 

A final comment concerns the range of validity of the thermal-mass description. The perturbative computation of thermal corrections is formally justified only when the relevant couplings are weak, such that $m_{\rm th}\sim gT \ll T$. From Eq.~(\ref{eq:th_masses}) one finds that the thermal masses of quarks and leptons satisfy this condition, supporting the perturbative treatment. The situation is less clear for gluons and $W$ bosons, whose thermal masses are of order $T$. In this regime the perturbative description becomes less reliable, and the physical thermal masses may differ appreciably from the perturbative estimates.

This uncertainty is nevertheless not expected to affect our conclusions significantly. 
Indeed we checked that the contribution of gluons to the friction depends only mildly on the precise value of $m_g$.
For the $W$ bosons, on the other hand, the large thermal mass strongly suppresses the out-of-equilibrium effects, rendering their contribution subdominant compared with that of the top quark. Moderate variations in the value of the $W$ thermal mass therefore do not alter the qualitative and quantitative results, as we have explicitly checked.

\subsection{Numerical implementation}

The brackets are computed numerically by discretising the momentum variable on a finite grid, following the approach of Refs.~\cite{DeCurtis:2023hil,DeCurtis:2024hvh}. We expand the relevant functions on the orthonormal basis ${e_m(q)}$, whose elements are non-vanishing only at the grid point $|\vec q\,|_m$. Using the scalar product defined in Eq.~(\ref{eq:sc_product}), one finds
\begin{equation}
e_m(q_n)=\delta_{mn}
\left[\frac{d|\vec q\,|_m}{dm}\frac{|\vec q\,|^2_m}{E_{q,m}} \,f_0(q_m)\right]^{-1/2}.  
\label{eq: basis}
\end{equation}

The grid consists of $M=100$ points spanning the range $\beta |\vec q\,| \le 20$. The points are distributed quadratically according to
\begin{equation}
\beta \,q_n = 20 \,(n/M)^2\,,
\end{equation}
providing a finer resolution in the low-momentum region where the kernels exhibit their most pronounced structure (see Figs.~\ref{fig: Gl top} and~\ref{fig: Gl W}) while keeping the evaluation of the brackets computationally efficient.

In the discretised basis, the kernel operators $\mathcal G_l$ are represented by $M\times M$ Hermitian matrices,
\begin{equation}
\mathcal U_{l,ab}=e_a\cdot \mathcal G_l\cdot e_b\,.
\end{equation}
Diagonalising these matrices yields the eigenvalues ${\lambda_{l,i}}$ and eigenvectors ${v_{l,i}}$, from which the corresponding eigenfunctions ${\zeta_{l,i}(q)}$ are obtained by interpolation.

The kernel functions $\mathcal Q$ and $\mathcal K$ (or equivalently $\mathcal G$) depend on the underlying model only through the scattering amplitudes entering the collision integrals. 
Consequently, the computation of $\mathcal Q$ and $\mathcal K$, as well as the diagonalisation of the latter, is performed only once for a given model.
The resulting spectral decomposition can then be reused throughout the numerical solution of the Boltzmann equations.

The strategy outlined above allows the collision integrals to be treated efficiently, reducing the solution of the Boltzmann equation to a sequence of one-dimensional integrations. This framework will be employed throughout the phenomenological analysis presented in the following sections.

\section{Analysis of the kernel functions}\label{sec:kernels}

We begin by analysing the impact of thermal masses on the kernel functions entering the Boltzmann equations. In the approximation adopted here, the kernels depend exclusively on Standard Model interactions. As a consequence, their determination can be performed once and for all, independently of the specific singlet-extended Higgs-sector scenario considered later on. The discussion in this section therefore focuses on general properties of the collision operators, which can be regarded as model-independent features of the thermal plasma.

Before presenting the numerical results, let us comment on an additional approximation adopted in the computation of the kernels. As discussed above, the annihilation and scattering processes entering the collision operator are evaluated using massless kinematics. Moreover, in the calculation of the source kernels $\mathcal Q$, we neglect the dependence of the top-quark and $W$-boson masses on the Higgs VEV and approximate their equilibrium distributions using only the corresponding thermal masses. With these assumptions, the $\mathcal Q$ kernels become independent of the wall coordinate $z$. The integral kernels $\mathcal K$, on the other hand, depend only on the equilibrium distributions of gluons and light quarks. Since the masses of these particles are dominated by thermal effects, their dependence on the Higgs VEV is negligible, and the corresponding kernels can likewise be treated as independent of $z$.

\subsection{The $\mathcal Q$ kernel functions}

\begin{figure}
\centering
\includegraphics[width=0.47\linewidth]{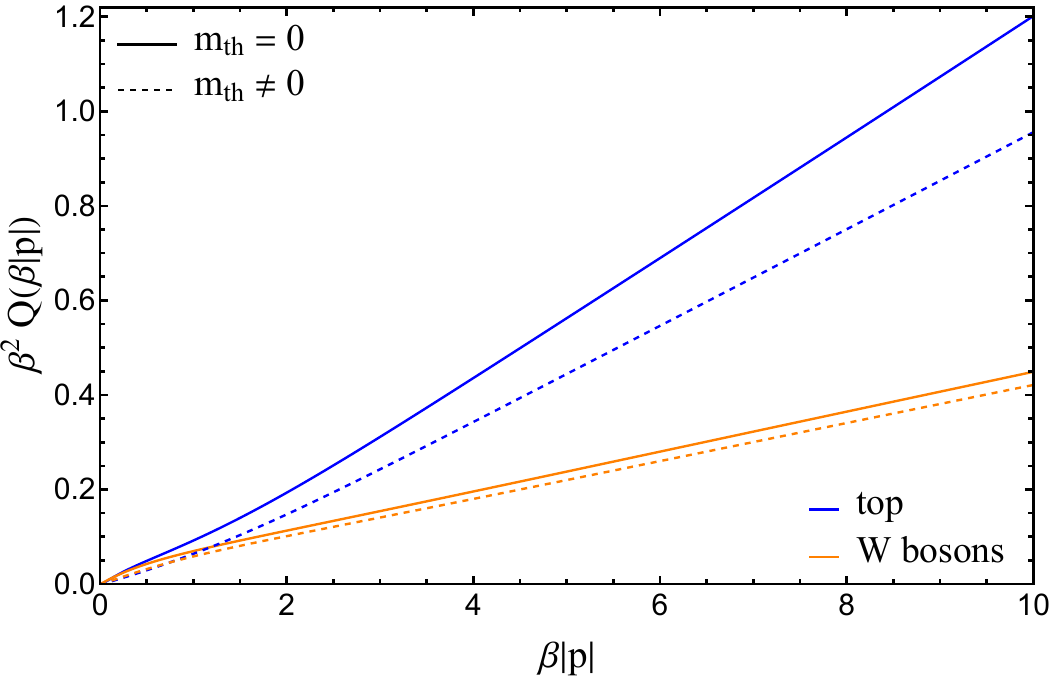}
\qquad
\includegraphics[width=0.47\linewidth]{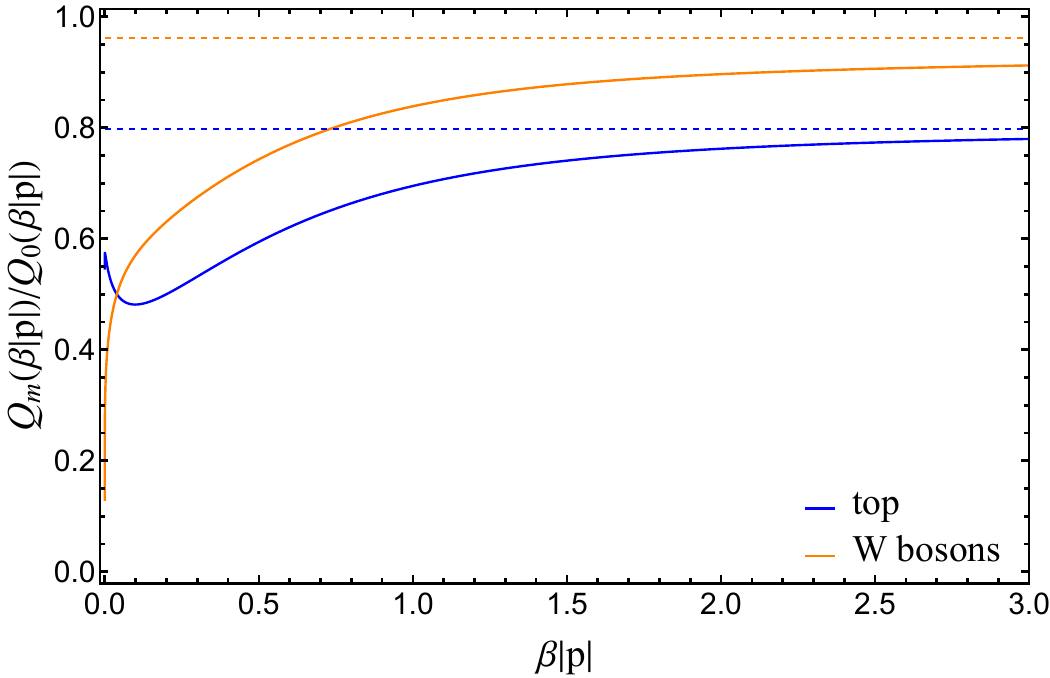}
\caption{\textit{Left panel}. Kernel function $\mathcal Q$ for the top (blue lines) and W bosons (orange lines) in the case where thermal masses are neglected (solid lines) and in the case where thermal masses are included in the collision operator (dashed lines). \textit{Right panel}. Ratio between the ``massive" and ``massless" kernels for the top (blue line) and W bosons (orange line). The blue and orange dashed lines show the asymptotic values of the ratio for large $\beta |\vec p\,|$. }
\label{fig: Kernel Q}
\end{figure}

We begin our numerical analysis by examining the impact of thermal masses on the kernel functions $\mathcal Q$. Figure~\ref{fig: Kernel Q} shows the results for the kernels $\mathcal Q_t$ and $\mathcal Q_W$. The left panel displays the kernels over the momentum range $\beta |\vec p\,| \le 10$, while the right panel shows the ratio between the kernels computed with and without thermal masses,
\begin{equation}
\frac{\mathcal Q_m(\beta |\vec p\,|)}{\mathcal Q_0(\beta |\vec p\,|)},
\end{equation}
for $\beta |\vec p\,| \le 3$. Here and throughout the paper, the labels `massive' and `massless' refer respectively to kernels computed with thermal masses included in the equilibrium distribution functions and to kernels obtained using massless distributions.

The figure reveals a qualitatively different behaviour for the two kernels. The inclusion of thermal masses leads to an $\mathcal O(1)$ suppression of the top kernel over the entire momentum range considered. In contrast, the $W$-boson kernel is significantly modified only in the infrared region $\beta |\vec p\,| \lesssim 1$, while it rapidly approaches the massless result at larger momenta.

The behaviour of the top kernel can be understood from the relative importance of the various scattering channels. At low momenta, $\mathcal Q_t$ is dominated by the annihilation process $t\bar t\to gg$ and by the contribution to $tg\to tg$ associated with quark exchange. The inclusion of thermal masses suppresses both channels by approximately a factor of two
visible in the right panel of Fig.~\ref{fig: Kernel Q}. The contribution to $tg\to tg$ associated with gluon exchange, although subdominant in this regime, receives a larger correction and is responsible for the non-trivial momentum dependence observed at very low momenta.

The situation changes at larger values of $\beta |\vec p\,|$. In this regime, the dominant contributions arise from the gluon-exchange component of $tg\to tg$ and from scattering with light quarks, $tq\to tq$. While the latter is only mildly affected by thermal masses, the former is again suppressed by roughly a factor of two. The asymptotic value $\mathcal Q_{t,m}/\mathcal Q_{t,0}\simeq 0.79$ therefore results from the interplay between these two channels and is largely driven by the inclusion of the gluon thermal mass.

The behaviour of the $W$-boson kernel is qualitatively different. At low momenta, $\mathcal Q_W$ is dominated by the processes $Wq\to qg$ and $Wl\to Wl$, while $WW\to l\bar l$ provides an additional contribution of roughly $10\%$ in the massive case. The inclusion of thermal masses strongly suppresses the first two channels, by approximately an order of magnitude as $\beta |\vec p\,|\to 0$, giving rise to the rapid drop visible in the right panel of Fig.~\ref{fig: Kernel Q}. These corrections are primarily driven by the gluon thermal mass in $Wq\to qg$ and by the $W$ thermal mass in $Wl\to Wl$. In contrast, the thermal masses of the fermions have a negligible impact on the kernel.

At large momenta, both the massless and massive kernels are dominated by the scattering process $Wf\to Wf$, which accounts for more than $80\%$ of the total contribution. Since this channel is only mildly affected by thermal masses, receiving a correction of about $4\%$, the ratio $\mathcal Q_{W,m}/\mathcal Q_{W,0}$ approaches the asymptotic value $0.96$. This limiting behaviour is therefore largely determined by the thermal correction to the $Wf\to Wf$ channel.

\subsection{The $\mathcal K$ integral kernels}

\begin{figure}[t]
    \centering
\includegraphics[width=0.47\linewidth]{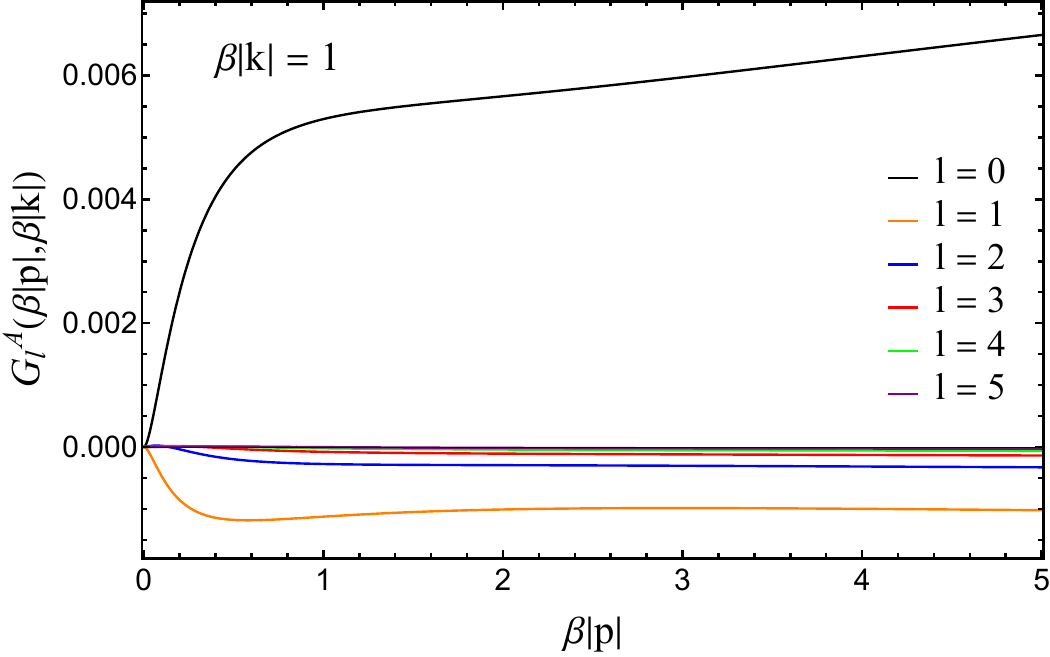}
\qquad
\includegraphics[width=0.47\linewidth]{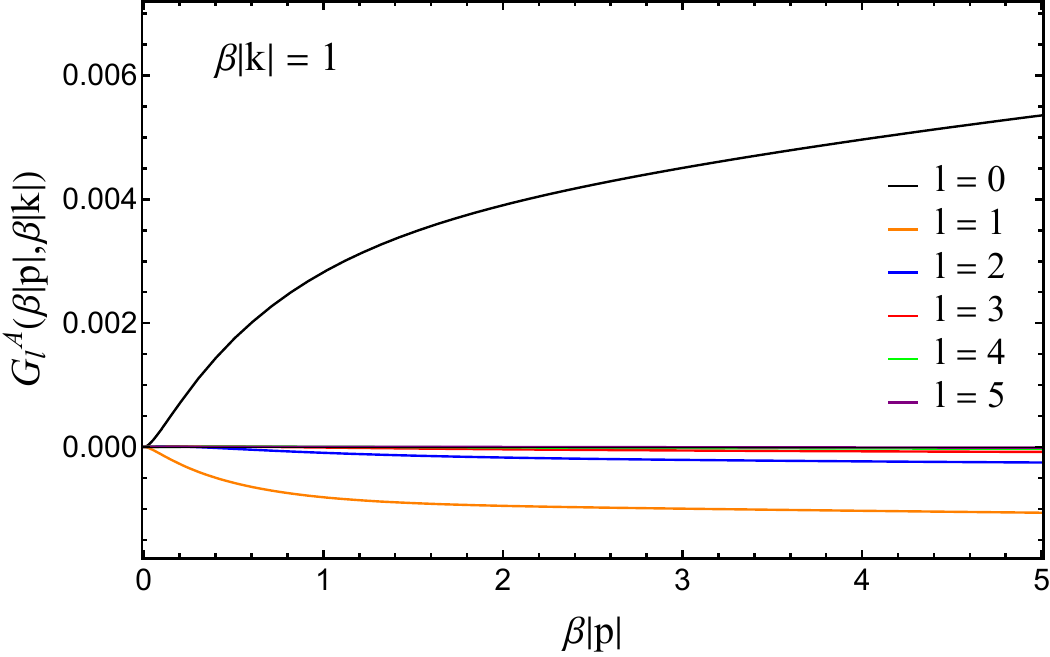}
\vskip 8pt
\includegraphics[width=0.47\linewidth]{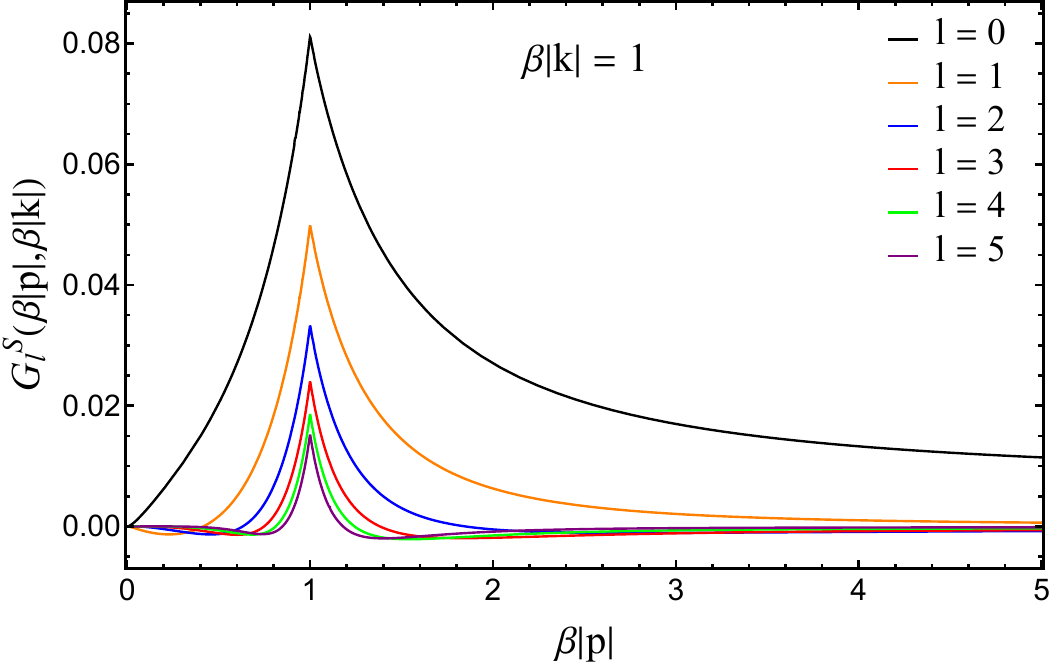}
\qquad
\includegraphics[width=0.47\linewidth]{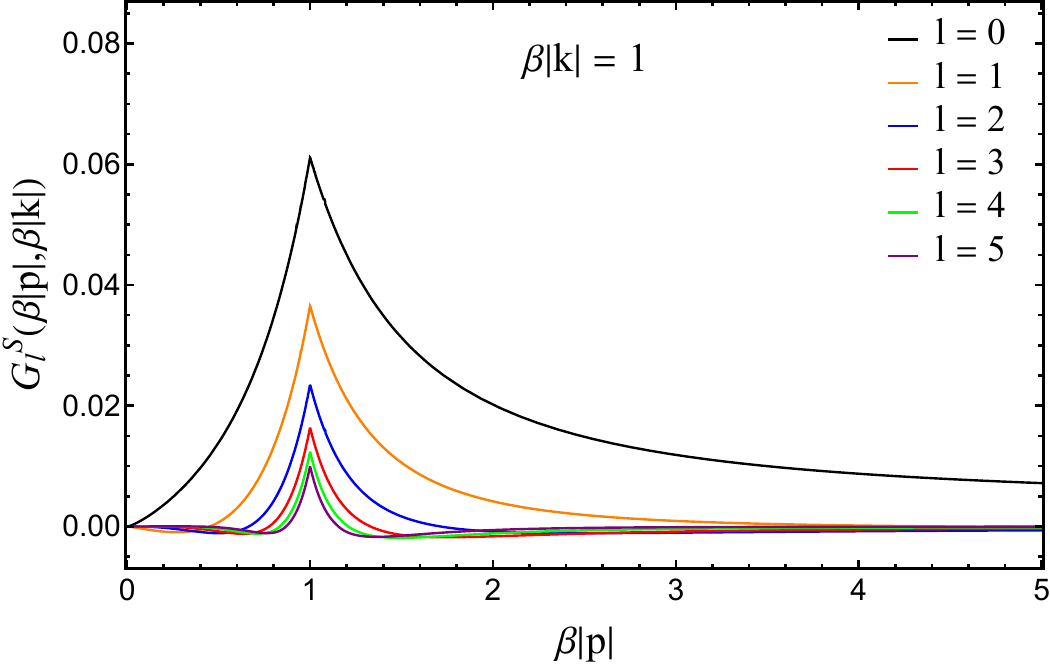}
    \caption{Liouville decomposition of the annihilation (first row) and scattering (second row) kernels for the top quark. The left column shows the kernels obtained with massless equilibrium distributions, while the right column includes thermal masses in the distribution functions.}
    \label{fig: Gl top}
\end{figure}

Moving now to the integral $\mathcal K$ kernels, Fig.~\ref{fig: Gl top} shows the annihilation and scattering kernels $\mathcal G_{l,pk}$ for the top quark, evaluated at $\beta |\vec k|=1$, for the first six Legendre blocks. The left and right panels correspond respectively to the massless and massive cases. The comparison between the two shows that thermal masses affect the annihilation and scattering kernels in slightly different ways. For the scattering kernels, the dominant effect is an overall reduction in magnitude, while their momentum dependence remains essentially unchanged. The annihilation kernels are also suppressed, but thermal masses induce a more noticeable modification of their low-momentum behaviour, smoothing the kink visible for $\beta |\vec p\,| \sim 0.5$ in the massless case.

When comparing the two cases, however, it is important to recall that thermal masses affect not only the kernels themselves but also the functional measure $\mathcal Dq$ appearing in Eq.~(\ref{eq: bracket decomposed}) and, consequently, the basis functions $e_m(q)$ defined in Eq.~(\ref{eq: basis}). The impact of thermal masses on the final brackets is therefore not fully captured by the visual comparison of the kernels alone.

The annihilation kernels are smooth and exhibit a strong hierarchy among Legendre modes, with the $l=3$ block already strongly suppressed. The scattering kernels show a markedly different behaviour: they develop a pronounced peak around $|\vec p\,|\simeq |\vec k|$, whose magnitude is roughly one order of magnitude larger than that of the corresponding annihilation kernel in the $l=0$ block, and decrease much more slowly with $l$. These features suggest, in agreement with the massless analysis of Ref.~\cite{DeCurtis:2024hvh}, that the scattering channels dominate the spectral decomposition. As a consequence, a relatively large number of Legendre blocks must be retained to accurately reconstruct the collision operator.

\begin{figure}
\centering
\includegraphics[width=0.47\linewidth]{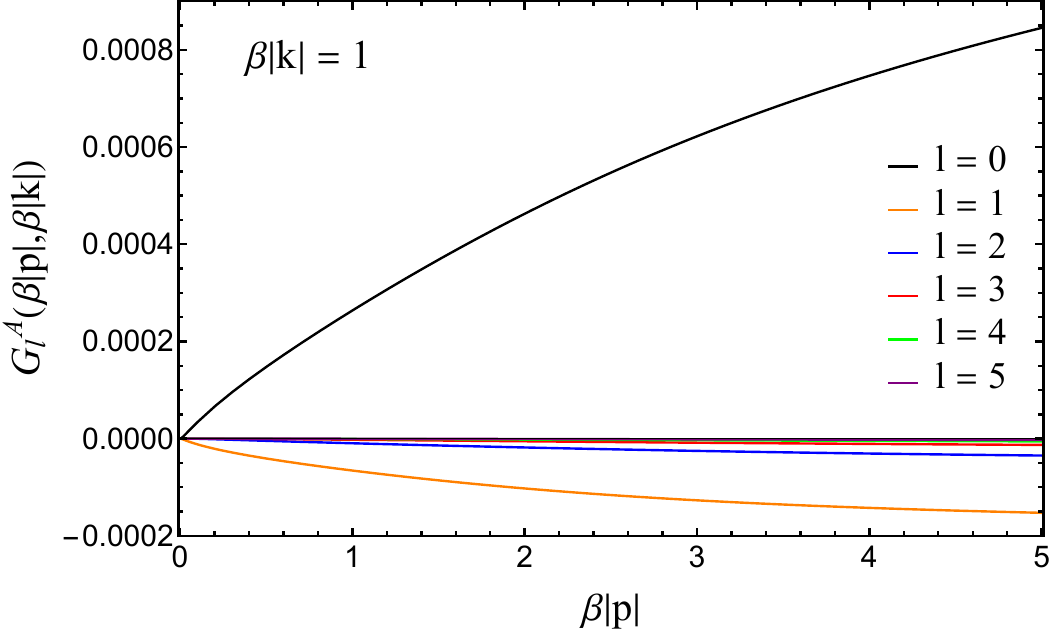}
    \qquad
\includegraphics[width=0.47\linewidth]{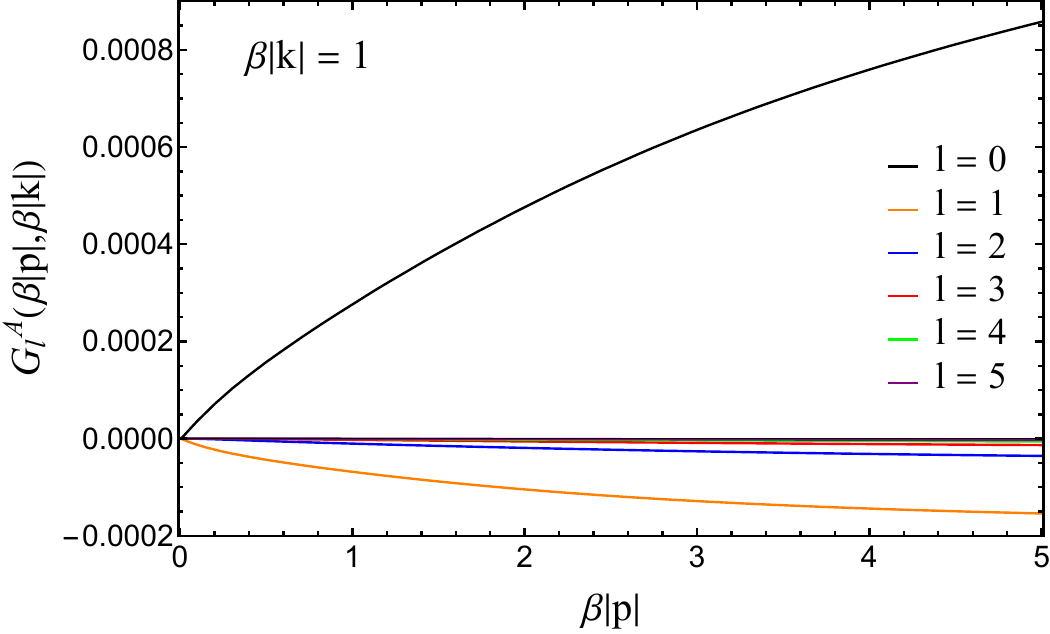}
\vskip 8pt
\includegraphics[width=0.46\linewidth]{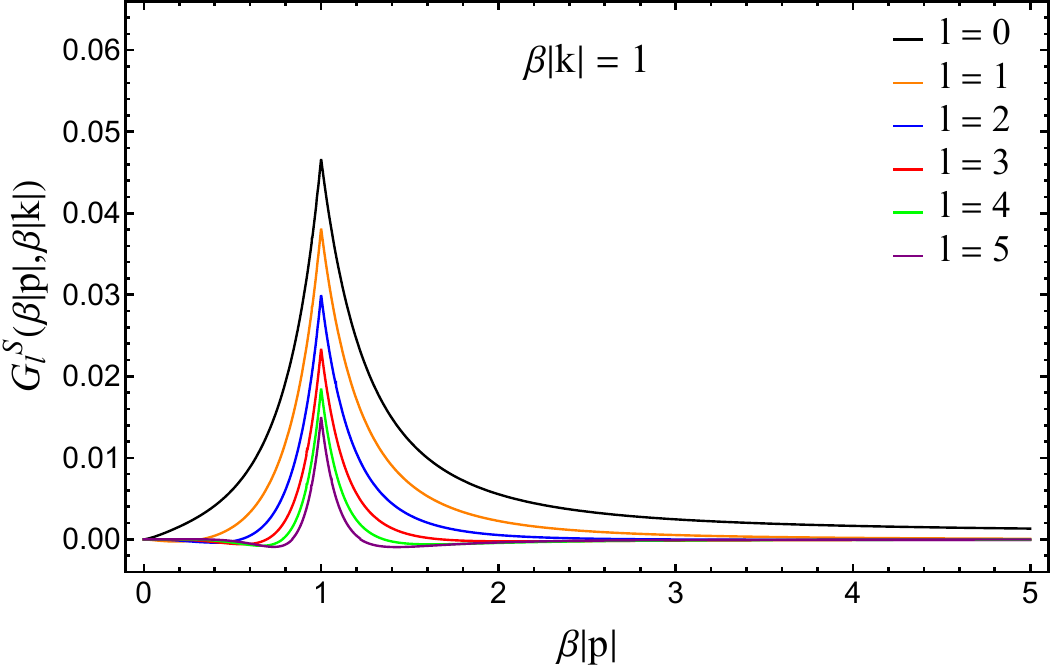}
\qquad\,
\includegraphics[width=0.46\linewidth]{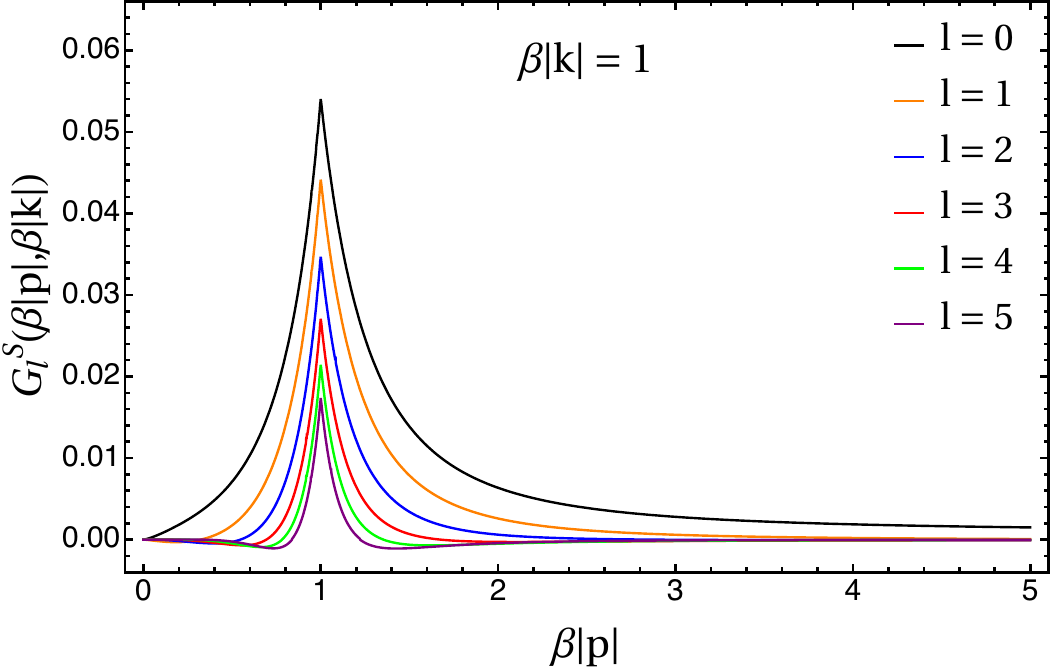}
    \caption{Liouville decomposition of the annihilation (first row) and scattering (second row) kernels for the W bosons. The panel layout is the same as in Fig.~\ref{fig: Gl top}.}
    \label{fig: Gl W}
\end{figure}

Similar observations can be made for the $W$-boson kernels $\mathcal G_l$ shown in Fig.~\ref{fig: Gl W}. In this case, however, the impact of thermal masses on the kernels is even milder. The annihilation kernels are nearly indistinguishable in the massless and massive cases, while the scattering kernels retain essentially the same momentum dependence and are only slightly enhanced when thermal masses are included. As in the top-quark case, the scattering kernels exhibit a much slower decrease with the Legendre block $l$ than the annihilation kernels.

The slight enhancement of the scattering kernels might appear surprising, given the suppression observed in the source kernel $\mathcal Q_W$. The origin of this behaviour can be traced back to the way thermal masses enter the equilibrium distribution functions appearing in the collision integrals. Since the dominant interactions of the $W$ bosons involve quarks and leptons, the equilibrium distribution functions entering $\mathcal G_l$ are predominantly fermionic. The enhancement visible in Fig.~\ref{fig: Gl W} is therefore driven by the corresponding Pauli-blocking factors. As discussed above, however, thermal masses also modify the functional measure and the basis functions entering the spectral decomposition. Together with the overall factor $f_0(p)$ multiplying $\langle \delta f\rangle$, these effects act in the opposite direction and ultimately lead to a suppression of the eigenvalues and of the corresponding brackets.

\begin{figure}
    \centering    \includegraphics[width=0.47\linewidth]{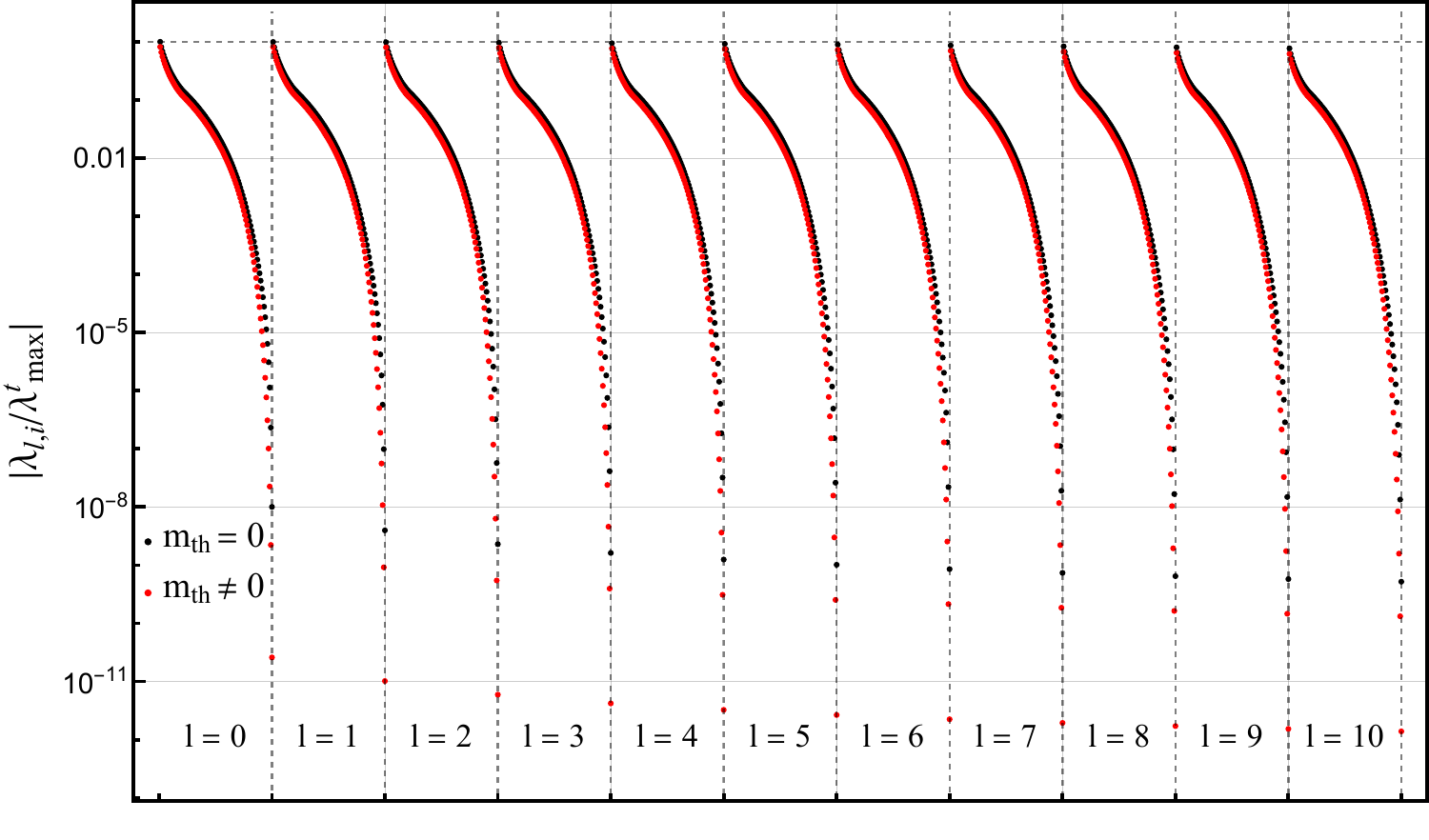}
    \qquad
    \includegraphics[width=0.47\linewidth]{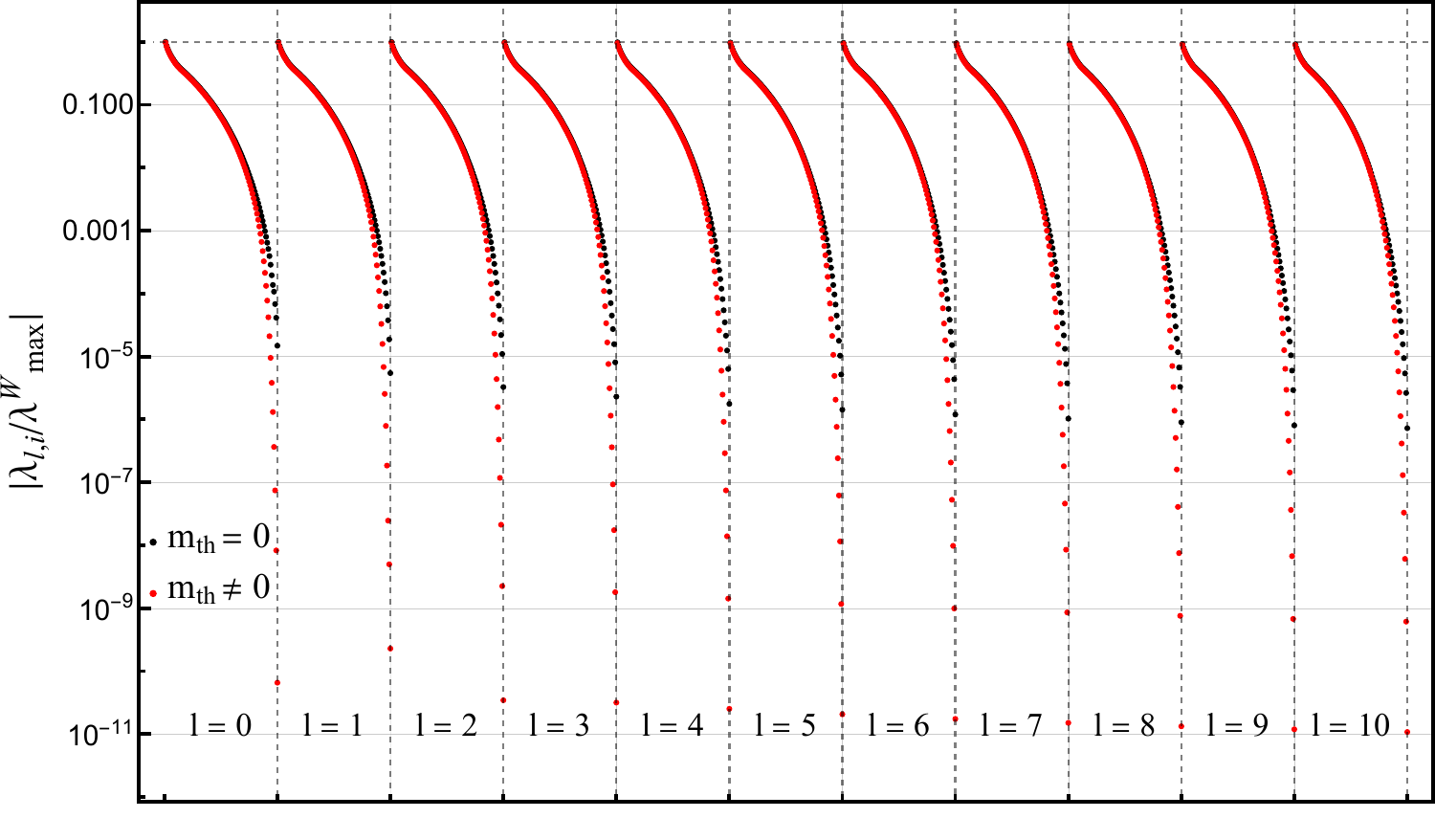}
    \caption{
    \textit{Left panel}. Eigenvalues of the top-quark kernel operator $\mathcal G_l^{t}$ for the first eleven Legendre blocks in the massless (black points) and massive (red points) cases. For each block, the first 100 eigenvalues $\lambda^{t}_{l,i}$, normalised to the largest one, $\lambda_{\rm max}^{t}\equiv\lambda^{t}_{0,0}$, are shown. \textit{Right panel}. Same as the left panel, but for the $W$-boson kernel operator $\mathcal G_l^{\scriptscriptstyle W}$, with eigenvalues normalised to $\lambda_{\rm max}^{W}\equiv\lambda^{\scriptscriptstyle W}_{0,0}$.
    }
    \label{fig: eigenvalues all}
\end{figure}

The features discussed above are reflected in the eigenvalue spectra of the discretised operators $\mathcal G_l^{tt}$ and $\mathcal G_l^{\scriptscriptstyle WW}$, shown in Fig.~\ref{fig: eigenvalues all} for the first eleven Legendre blocks. The slow decrease of the eigenvalues with $l$ confirms that a relatively large number of Legendre blocks must be retained in order to accurately reconstruct the collision operator. Moreover, separating the annihilation and scattering contributions, one finds that the latter dominate the spectrum, while the annihilation eigenvalues are rapidly suppressed at increasing $l$.

The figure also illustrates the impact of thermal masses. In both the top and $W$ sectors, the eigenvalues obtained with thermal masses are smaller in magnitude than their massless counterparts, although the effect is considerably more pronounced for the top quark. In the $W$ case, this behaviour should be contrasted with the slight enhancement observed directly at the level of the kernels $\mathcal G_l$. As discussed above, the suppression of the eigenvalues originates from the thermal-mass dependence of the functional measure and basis functions entering the spectral decomposition.

\begin{figure}[t]
\centering
\includegraphics[width=0.47\linewidth]{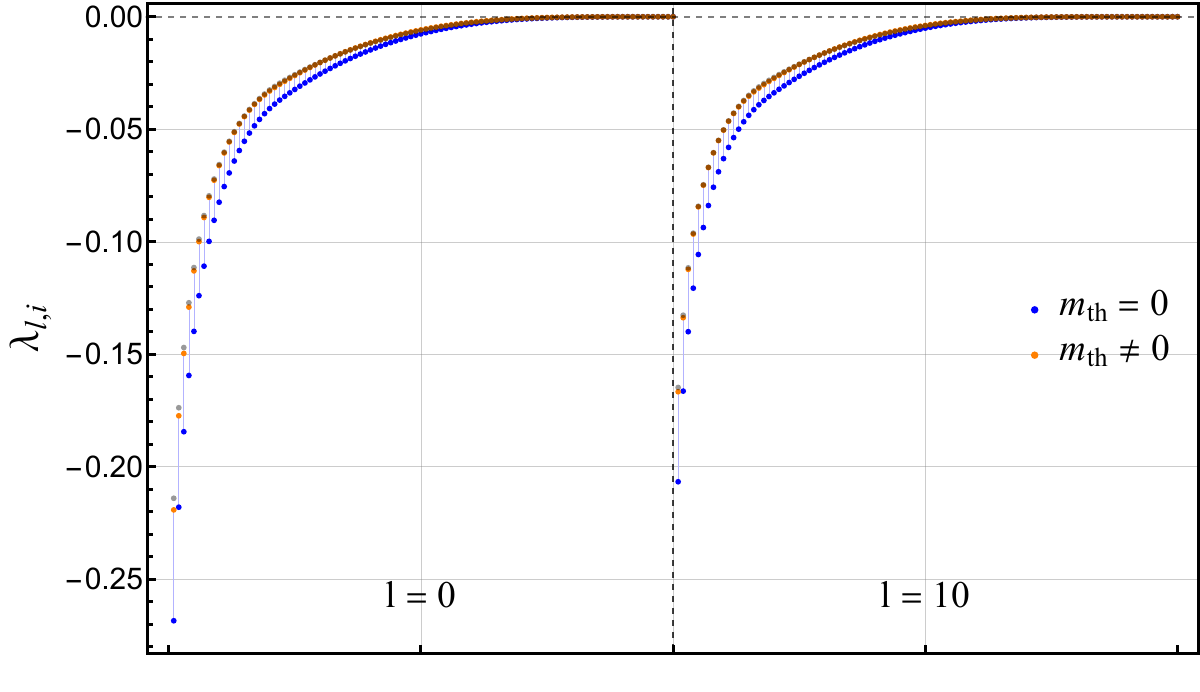}
\qquad \includegraphics[width=0.47\linewidth]{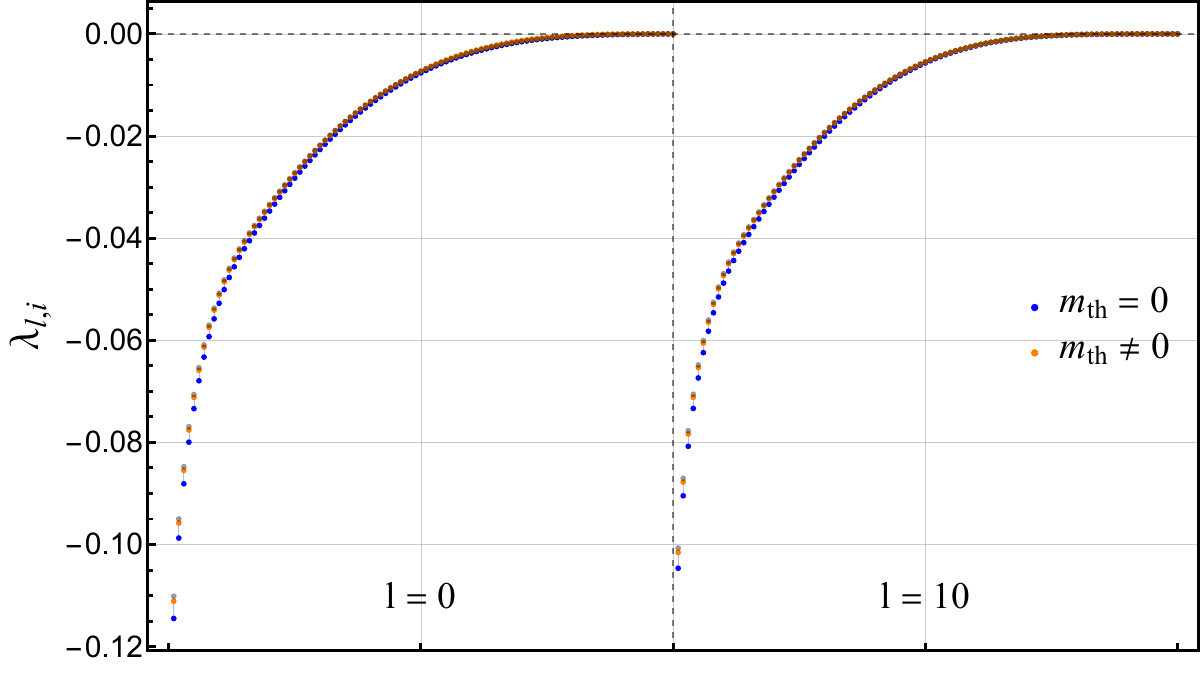}    
\caption{\textit{Left panel}. Comparison between the first 100 eigenvalues of the massless (blue points) and massive (orange points) top-quark kernel operator $\mathcal G_l^t$ for the first ($l=0$) and last ($l=10$) Legendre blocks included in the computation. Gray points denote the massless eigenvalues rescaled by the factor $\mathcal Q_{t,m}^+/\mathcal Q_{t,0}^+$, corresponding to the ratio between the asymptotic values of the kernel function $\mathcal Q_t$ in the massive and massless cases.
\textit{Right panel}. Same as the left panel, but for the $W$-boson kernel operator $\mathcal G_l^{\scriptscriptstyle W}$ and using the corresponding rescaling factor $\mathcal Q_{W,m}^{+}/\mathcal Q_{W,0}^{+}$.}
\label{fig: eigenvalues m-comparison}
\end{figure}

Figure~\ref{fig: eigenvalues m-comparison} provides a more detailed comparison between the massless and massive spectra. A striking feature is that, to a good approximation, the inclusion of thermal masses acts as an overall rescaling of the eigenvalues. For the top kernel, the rescaling factor closely matches the asymptotic ratio of the $\mathcal Q$ kernel functions, $\mathcal Q^+_{t,m}/\mathcal Q^+_{t,0}\simeq 0.79$, where $\mathcal Q^+$ denotes the large-momentum limit of $\mathcal Q$. A similar behaviour is observed for the $W$ bosons, although the agreement with the corresponding ratio $\mathcal Q^+_{W,m}/\mathcal Q^+_{W,0}$ is slightly less accurate, presumably due to the residual dependence of the eigenvectors and eigenfunctions on the equilibrium distributions. The figure also makes it apparent that the $W$-boson spectrum decreases more slowly with the Legendre block than the top-quark one.

\begin{figure}
    \centering
    \includegraphics[width=0.45\linewidth]{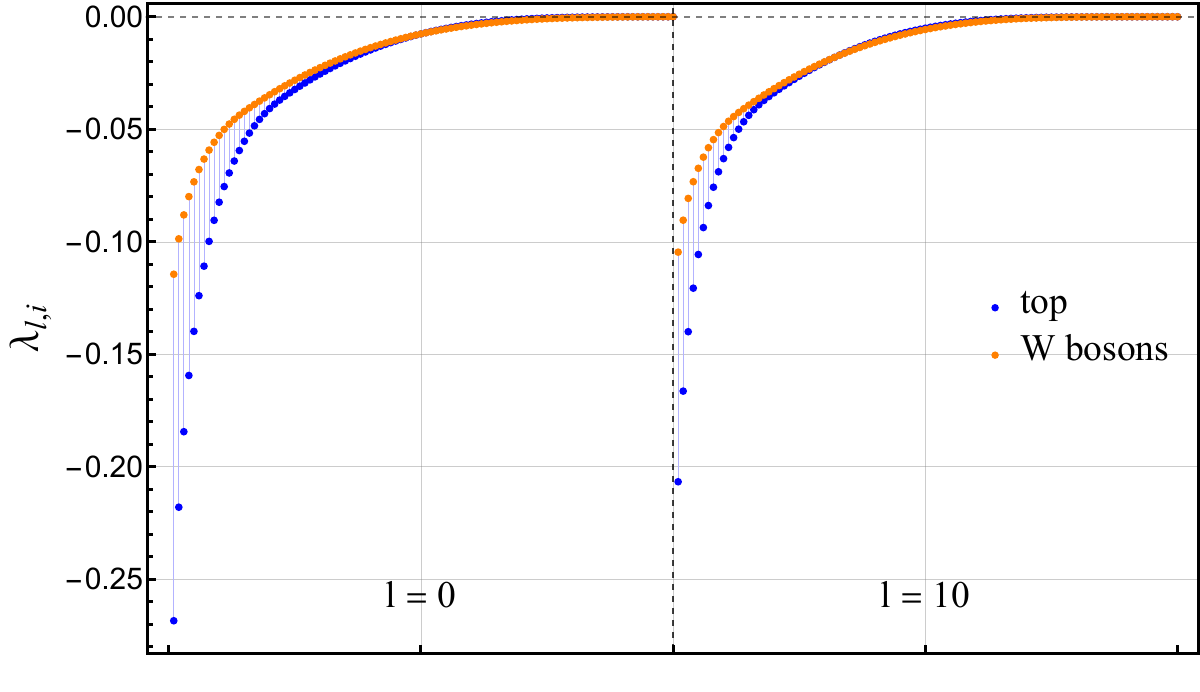}
    \qquad
    \includegraphics[width=0.45\linewidth]{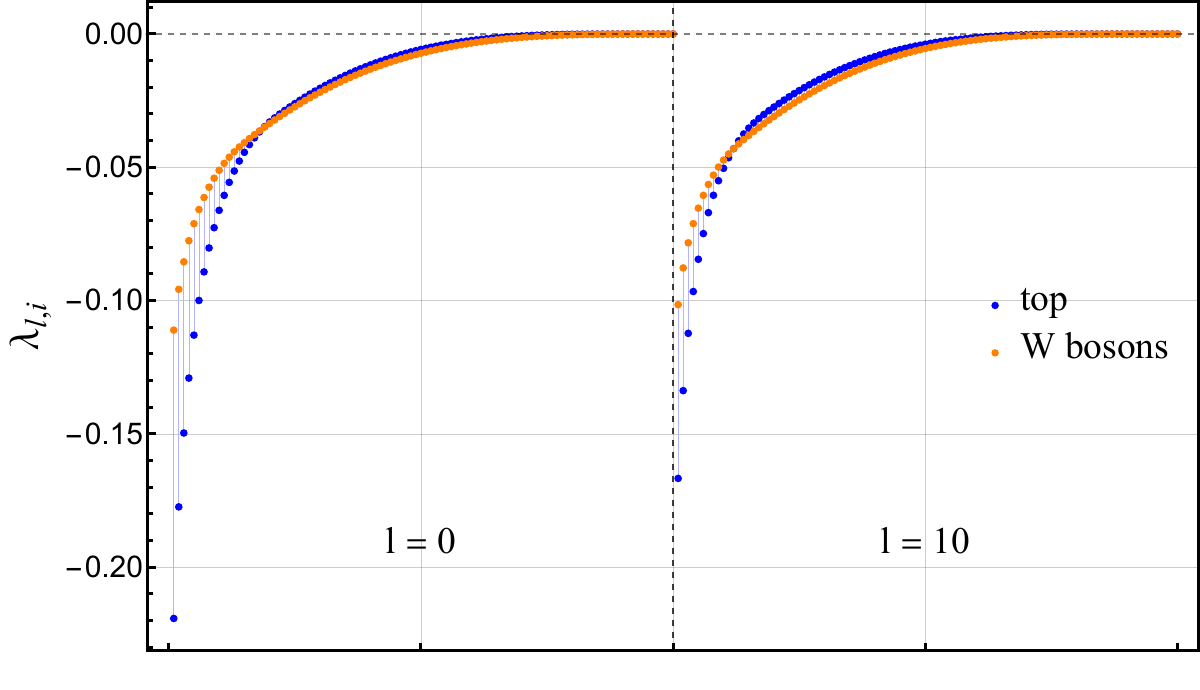}
    \caption{
    \textit{Left panel}. Comparison between the first 100 eigenvalues of the top-quark and $W$-boson kernel operators, $\mathcal G_l^{t}$ (blue points) and $\mathcal G_l^{\scriptscriptstyle W}$ (orange points), for the first ($l=0$) and last ($l=10$) Legendre blocks in the massless case. \textit{Right panel}. Same as the left panel, but for the massive case.
    }
    \label{fig: eigenvalues top vs W}
\end{figure}

Figure~\ref{fig: eigenvalues top vs W} compares the eigenvalue spectra of the top and $W$ kernel functions. In the massless case, the leading top-quark eigenvalues are typically larger than the corresponding $W$-boson ones by a factor of about two. The difference becomes less pronounced when thermal masses are included, and, for sufficiently large mode number, one even finds $\lambda^{\scriptscriptstyle W}_{l,i}>\lambda^{t}_{l,i}$.

The comparison of the spectra alone is, however, insufficient to determine the relative importance of the top and $W$ contributions to the friction. The final result depends on the interplay between the integral kernels $\mathcal K$, with the corresponding brackets, the kernel functions $\mathcal Q$, and the source terms $\mathcal S$, all of which are affected by thermal masses. This interplay will be analysed in the next section.

\section{Application to the singlet-extended Standard Model}\label{sec:application}

We now turn to the impact of thermal masses on the dynamics of bubble walls during a first-order phase transition. As a concrete application, we consider the $\mathbb Z_2$-symmetric singlet extension of the Standard Model (SSM), which provides one of the simplest frameworks capable of realising a two-step phase transition. In this scenario, the first transition generates a VEV for the singlet field, while the second step can induce a strongly first-order electroweak phase transition.

The model and the theoretical framework used to describe the electroweak phase transition, including the dynamics of both the scalar fields and the plasma, have been presented in detail in previous studies. Here we restrict ourselves to a brief summary of the equations relevant for the determination of the bubble-wall properties and refer the reader to Ref.~\cite{Branchina:2025jou} for a comprehensive discussion.

The SSM enlarges the scalar sector of the Standard Model by the addition of a real scalar field $s$, with tree-level potential 
\begin{equation}
	V_0(h,s) = \frac{\mu_h^2}{2}  h^2 + \frac{\lambda_h}{4}  h^4 + \frac{\mu^2_s}{2} s^2 + \frac{\lambda_s}{4} s^4 + \frac{\lambda_{hs}}{2} h^2 s^2. 
\label{eq: V0 SSM}
\end{equation}
The extended scalar sector includes three additional free parameters, that can be chosen to be the self-coupling $\lambda_s$, the portal coupling to the Higgs boson $\lambda_{hs}$, and the physical mass $m_s$ in the electroweak vacuum $(h,s)=(v,0)$, which is obtained as $m_s^2=\mu_s^2 +\lambda_{hs}  v^2$. As for the determination of the one-loop contribution, we resort to an on-shell renormalisation and the Parwani resummation scheme, as detailed in~\cite{Branchina:2025jou}. 

The equations of motion describing the dynamics of the scalar fields read\footnote{The additional contributions beyond the WKB approximation of Refs.~\cite{Ai:2025bjw,Ramsey-Musolf:2025jyk} are only relevant to determine the friction in the $v_w\to 1$ limit.}~\cite{Moore:1995ua,Moore:1995si}
\begin{align}
	E_h \equiv - \partial^2_z h + \frac{\partial V(h,s,T)}{\partial h} + F_h^{\scriptscriptstyle \rm OOE}(z) = 0, \\
	E_s \equiv - \partial^2_z s + \frac{\partial V (h,s,T)}{\partial s} +  F_s^{\scriptscriptstyle \rm OOE}(z) =0,
\label{eq: scalar eq}
\end{align} 
where $V$ is the effective potential and $F_{j}^{\scriptscriptstyle \rm OOE}$ ($j=h,s$) indicates the out-of-equilibrium (OOE) contribution arising from the coupling of the $i$-th plasma species to the wall ($\vec \phi = \{h,s\}$), 
\begin{equation}
	F^{\scriptscriptstyle \rm OOE}_j(z)=\sum_i \frac{n_i}{2}\frac{\partial m^2_i}{\partial \phi_j}\int\frac{d^3p}{(2\pi)^3 E_p} \,\delta f_i.
\label{eq: F def}
\end{equation}
These equations can be approximately solved by assuming a $\tanh$ profile for the scalar fields:
\begin{align}
	h(z) &= \frac{h_{-}}{2} \left( 1 + \tanh{\left( \frac{z}{L_h}\right)}\right),  \nonumber \\
	s(z) &= \frac{s_{+}}{2} \left( 1 - \tanh{\left( \frac{z}{L_s} - \delta_s \right)}\right),
	\label{eq: ansatz scalars}
\end{align}
where 
$h_{-}$ and $s_{+}$ are the VEVs on the two sides of the wall, i.e.~in the unbroken and broken phase respectively,
\begin{equation}
	\frac{\partial V(h_{-}, 0, T_-)}{\partial h}=0, \qquad \qquad \frac{\partial V(s_{+}, 0, T_+)}{\partial s}=0.
\end{equation}
The $\tanh$ ansatz allows one to recast the dynamic equations~\eqref{eq: scalar eq} as four integral constraints:
\begin{align}
	P_h &= \int_{-\infty}^{\infty} dz \,E_h h' = 0, \qquad \qquad  G_h = \int_{-\infty}^{\infty} dz \,E_h \left(2\frac{h}{h_{-}} -1\right) h' = 0, \nonumber \\
	P_s &= \int_{-\infty}^{\infty} dz \,E_s s' = 0, \qquad \qquad  G_s = \int_{-\infty}^{\infty} dz\, E_s \left(  2\frac{s}{s_{+}} -1\right) s' = 0.
\label{eq: constraints}
\end{align}
Solving these constraints determines the profile parameters $\delta_s,\, L_h,\, L_s$ and the wall velocity $v_w$.\footnote{$v_w$ implicitly enters in the profiles through $T_-$, which is unknown regardless of the combustion regime, and to which $v_w$ is related through the fluid equations.}

The equations~\eqref{eq: constraints} must be solved together with the Boltzmann equation, for the top and the $W$ bosons, and the fluid equations obtained from the conservation of the stress-energy tensor (the sum extends to all the species in the plasma)
\begin{equation}
T_{i, \mu\nu}=\sum_i\int\frac{d^3q}{(2\pi)^3 E_q}q_\mu q_\nu f_i(q,x), 
\label{eq: stress energy tensor}
\end{equation}
in the $z$-direction, $\partial^z T_{z0}= \partial^z T_{zz} = 0$, from which
\begin{align}
T_{30} \equiv w\, \gamma^2 v_p + T_{30}^{\scriptscriptstyle \rm OOE} = c_1, \hspace{3cm}
\label{eq: hydro eq. 1}
\\
T_{33} \equiv \sum_{j=\{h,s\}}\frac{(\partial_z \phi_j)^2}{2} - V(h,s, T) + w\, \gamma^2 v_p^2 + T_{33}^{\scriptscriptstyle \rm OOE} = c_2. 
\label{eq: hydro eq. 2}
\end{align}
In the equations above, $w=-T\partial_T V$ denotes the enthalpy density, $v_p$ is the plasma velocity and $\gamma$ its Lorentz factor, $T^{\scriptscriptstyle \rm OOE}_{\mu\nu}$ represents the out-of-equilibrium contribution to Eq.~\eqref{eq: stress energy tensor}, and $c_{1,2}$ are integration constants determined by the asymptotic values of the plasma temperature and velocity, $T_\pm$ and $v_\pm$. Here the $+$ ($-$) subscript denotes quantities in front of (behind) the wall.

\subsection{Analysis of two benchmarks}\label{sec:benchmarks}

We begin our discussion by considering two benchmark points. Besides providing concrete examples, they allow us to analyse in detail how thermal masses modify the out-of-equilibrium friction and to disentangle the effects associated with the different terms entering the Boltzmann equation.

The benchmarks, denoted by BP$1$ and BP$2$, correspond to the parameter choices
\begin{equation}
\begin{array}{l}
\textrm{BP}1 :\quad (m_s,\lambda_{hs},\lambda_s ) = (79\, {\rm GeV},\, 0.39, 1)\,,\\
\rule{0pt}{1.25em}\textrm{BP}2 :\quad (m_s,\lambda_{hs},\lambda_s ) = (100\, {\rm GeV},\, 0.39, 1)\,.
\end{array}
\end{equation}
The two points are broadly comparable to the benchmarks analysed in Refs.~\cite{DeCurtis:2023hil,DeCurtis:2024hvh}\footnote{Notice that in Refs.~\cite{DeCurtis:2023hil,DeCurtis:2024hvh} a different normalisation for the coupling $\lambda_{hs}$ was adopted, as well as a different renormalisation scheme.}. They were chosen because they are representative of the typical size of the out-of-equilibrium corrections found across the parameter space~\cite{Branchina:2025adj}.

\begin{figure}
\centering
\includegraphics[width=0.75\linewidth]{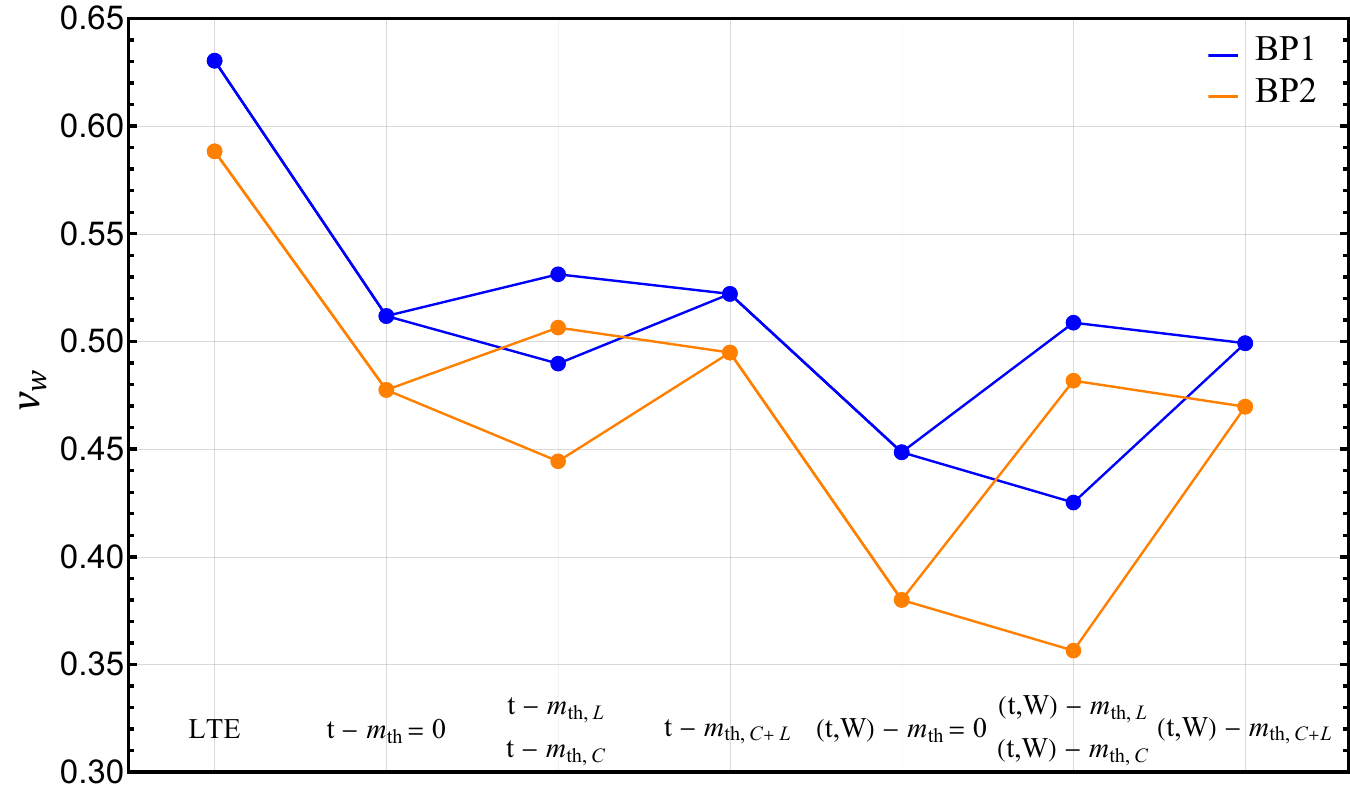}
\caption{Wall velocity for the benchmark points BP$1$ and BP$2$, for the different treatments of thermal masses discussed in the text. The leftmost point corresponds to the LTE solution. For the remaining points, the label before the dash identifies the out-of-equilibrium species ($t$ for the top quark and $(t,W)$ for the top quark plus $W$ bosons), while the label after the dash specifies the treatment of thermal masses: $m_{\rm th}=0$ (massless approximation), $C$ (collision integrals only), $L$ (Liouville operator only), and $C+L$ (both). }
\label{fig: vw benchmarks}
\end{figure}

Figure~\ref{fig: vw benchmarks} summarises the wall velocities obtained for the different treatments of thermal masses considered in this work: (i) local thermal equilibrium (LTE); (ii)-(iv) only top-quark out-of-equilibrium contributions, in the massless approximation, with thermal masses included only in the collision integrals, only in the Liouville operator, or in both; and (v)-(vii) the corresponding results obtained after including the $W$-boson contributions. The values of the remaining profile parameters, $\delta_s$, $L_h$ and $L_s$, together with the asymptotic temperatures $T_\pm$, are reported in Table~\ref{tab: results BP}.

\begin{table}[t]
\centering
{\small
\begin{tabular}{@{}c|c|c|c|c|c|c@{}}
BP1 & $v_w$ & $\delta_s$ & $L_h T_n$ & $L_s T_n$ & $T_+$ [GeV] & $T_-$ [GeV] \\
        \hline
\rule{0pt}{1.1em} LTE & 0.63 & 0.80 & 4.26 &3.71 & 103.91 & 98.41 \\ 
\rule{0pt}{1.1em}$m_{\rm th}=0$ & 0.45\ \  (0.51) & 0.79\ \ (0.79) & 4.60\ \ (4.67) & 3.78\ \ (3.82) & 95.04\ \ (96.39) & 93.24\ \  (93.25)\\ 
\rule{0pt}{1.em}$m_{\rm th,\,C}$ & 0.43\ \  (0.49) & 0.79\ \ (0.79) & 4.61\ \ (4.67) & 3.78\ \ (3.82) & 94.74\ \ (95.78) & 93.21\ \ (93.22) \\ 
\rule{0pt}{1.1em}$m_{\rm th,\, L}$ & 0.51\ \  (0.53) & 0.79\ \ (0.79) & 4.79\ \ (4.71) & 3.89\ \ (3.83) & 96.34\ \ (97.11) & 93.32\ \ (93.27)\\ 
\rule{0pt}{1.em}$m_{\rm th,\,C+L}$& 0.50\ \  (0.52) & 0.79\ \ (0.79) & 4.81\ \ (4.75) & 3.91\ \ (3.87) & 96.06\ \ (96.75) & 93.30\ \ (93.26)
\end{tabular}
\vskip 10pt 
\begin{tabular}{@{}c|c|c|c|c|c|c@{}}
BP2 & $v_w$ & $\delta_s$ & $L_h T_n$ & $L_s T_n$ & $T_+$ [GeV] & $T_-$ [GeV] \\
        \hline
\rule{0pt}{1.1em} LTE & 0.59 & 0.72 & 5.86 & 4.82 & 126.02 & 121.01 \\ 
\rule{0pt}{1.1em}$m_{\rm th}=0$ & 0.38\ \  (0.48) & 0.73\ \ (0.73) & 6.53\ \ (6.78) & 5.12\ \ (5.26) & 120.05\ \ (120.81) & 119.54\ \  (119.57)\\ 
\rule{0pt}{1.em}$m_{\rm th,\, C}$ & 0.36\ \  (0.44) & 0.73\ \ (0.73) & 6.56\ \ (6.75) & 5.13\ \ (5.24) & 119.96\ \ (120.44) & 119.53\ \ (119.56) \\ 
\rule{0pt}{1.1em}$m_{\rm th,\, L}$ & 0.48\ \  (0.51) & 0.73\ \ (0.73) & 6.93\ \ (6.83) & 5.33\ \ (5.28) & 120.88\ \ (121.33) & 119.59\ \ (119.58)\\ 
\rule{0pt}{1.em}$m_{\rm th,\, C+L}$& 0.47\ \  (0.49) & 0.73\ \ (0.73) & 6.93\ \ (6.87) & 5.34\ \ (5.30) & 120.71\ \ (121.09) & 119.58\ \ (119.58)
\end{tabular}
}
\caption{Results for the wall velocity $v_w$, the profile parameters $\delta_s$, $L_h$, and $L_s$, and the asymptotic temperatures $T_\pm$ for the benchmark points BP$1$ ($T_n=93.23\,\mathrm{GeV}$) and BP$2$ ($T_n=119.50\,\mathrm{GeV}$). Values in parentheses correspond to the case where only top-quark out-of-equilibrium contributions are included.}
\label{tab: results BP}
\end{table}

Several features are immediately apparent. First, out-of-equilibrium effects substantially reduce the wall velocity with respect to the LTE solution. Second, in the massless approximation the contribution of the $W$ bosons is sizeable, generating an additional reduction of the wall velocity comparable to that produced by the top quark. Finally, the inclusion of thermal masses significantly weakens the impact of the $W$ bosons, while producing only moderate corrections to the top-quark contribution.

The reduction of the $W$ boson contributions is particularly important because the massless approximation receives a substantial contribution from the infrared dynamics, where the validity of the Boltzmann description is questionable. Before analysing the microscopic origin of this behaviour, we first discuss how thermal masses affect the various terms entering the Boltzmann equation.

The decrease of $v_w$ observed when thermal masses are included only in the collision integrals can be understood as a consequence of the reduced interaction rates in the plasma. Thermal masses suppress the collision operator, making the relaxation towards equilibrium less efficient. As a result, larger deviations from equilibrium develop close to the wall, increasing the friction exerted on it and leading to a lower wall velocity.

A qualitatively different effect is observed when thermal masses are included in the Liouville operator. In this case, the source term $\mathcal S$ responsible for generating the out-of-equilibrium perturbations is reduced. This effect is present for both the top quark and the $W$ bosons, but it mainly affects low-energy modes with $|\vec p\,| \lesssim m_{\rm th}$. For the top quark, the resulting modification is relatively mild, since the thermal mass is small and, most importantly, it only moderately alters the Fermi-Dirac distribution. In contrast, the effect is much more pronounced for the $W$ bosons. In fact, thermal masses regulate the infrared Bose enhancement present in the massless approximation, strongly suppressing the contribution from soft gauge bosons. As a consequence, the out-of-equilibrium perturbations are reduced, leading to a smaller friction and therefore to a larger wall velocity. We note that the reduction of the friction induced by thermal masses in the Liouville operator agrees with the findings of Ref.~\cite{vandeVis:2025plm}, which adopted the same treatment of thermal effects.

\begin{figure}
    \centering
\includegraphics[width=.95\linewidth]{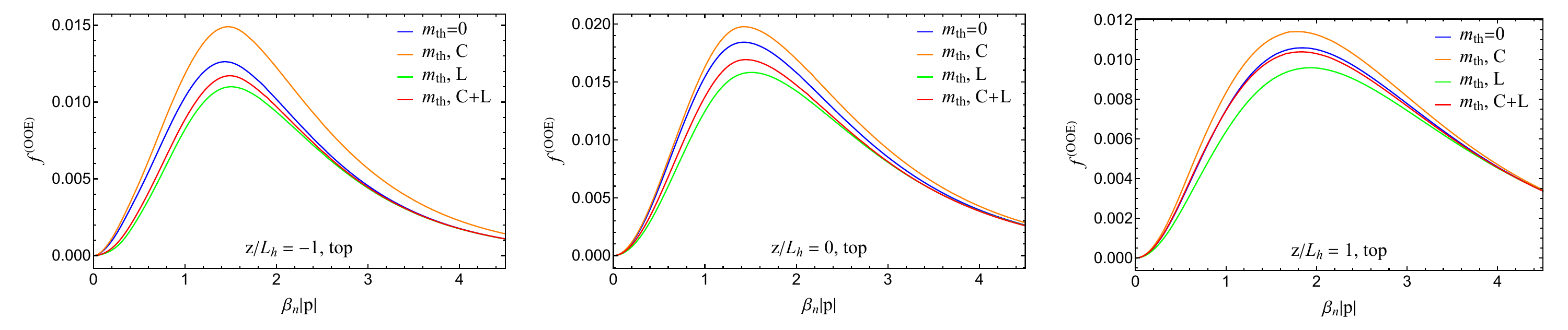}
\includegraphics[width=.95\linewidth]{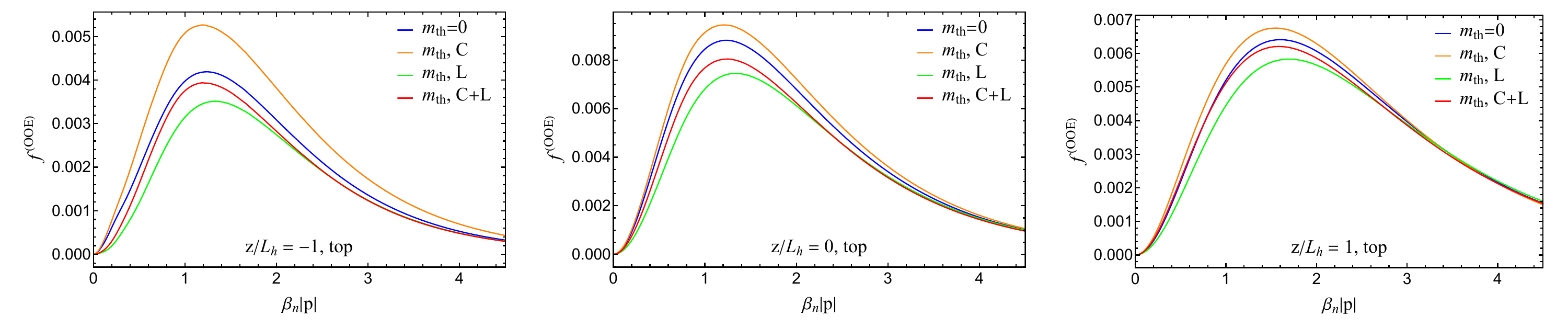} \caption{Top-quark friction integrand $f_t^{\scriptscriptstyle\rm (OOE)}$ for BP$1$ (upper row) and BP$2$ (lower row), evaluated on the solutions obtained including only top-quark out-of-equilibrium contributions. The panels correspond to positions in front of the wall ($z/L_h=-1$, left), at the wall centre ($z/L_h=0$, centre), and behind the wall ($z/L_h=1$, right).}
\label{fig: friction integrand only top}
\end{figure}

For the top quark, it turns out that the modifications of the collision integral and of the source term largely compensate each other, resulting in only a mild reduction of the overall out-of-equilibrium friction. The net effect of including thermal masses is an increase of the wall velocity of about $2\%$. The situation is markedly different for the $W$ bosons. In this case, in fact, the suppression of the source term greatly dominates over the reduction of the collision rate, leading to a substantial decrease of the friction they generate. As a result, the $W$-boson contribution becomes subdominant with respect to the top-quark one, and in the combined scenario with thermal masses in both $\mathcal L$ and $\mathcal C$ its inclusion lowers the wall velocity by only $\sim 4\%$ relative to the top-only approximation. Remarkably, the final result obtained after consistently including thermal masses is very close to that found in the much simpler approximation where only the top-quark contribution to the out-of-equilibrium friction is retained and all species are treated as massless.

To better understand the impact of thermal masses on the wall dynamics, it is useful to analyse the momentum dependence of the friction integrands entering the out-of-equilibrium force. In Fig.~\ref{fig: friction integrand only top} we show the contribution generated by the top quark
\begin{equation}
    f_t^{\scriptscriptstyle \rm (OOE)} (\vec p,z) = \frac{n_t}{2E_p}\frac{1}{h}\frac{\partial m^2_t}{\partial h}\delta f_t = \frac{n_t \,y_t^2}{2E_p}\delta f_t 
\label{eq: f_t OOE}
\end{equation}
for the two benchmark points. As we did before, we compare the results obtained including the thermal masses in different terms of the Boltzmann equation. The function $f_t^{\scriptscriptstyle \rm (OOE)}$ is plotted against $\beta_n |\vec p\,|$ for the three position values $z/L_h=-1,0,1$, which corresponds to locations in front, at the centre and behind the wall respectively. The friction is evaluated on the solution found by including only the top quark out-of-equilibrium contributions. 

In agreement with the previous discussion, we find that $f_t^{\scriptscriptstyle \rm (OOE)}$ bears only a mild dependence on the thermal masses. The inclusion of thermal masses in the collision integrals (orange curves) leads to an overall increase of the friction with respect to the case with no thermal masses (blue curves). The effect is somewhat larger outside the bubble, where all plasma species are massless, apart from thermal effects. The inclusion of thermal masses in the Liouville operator (green curves), instead, has an opposite impact as it induces a decrease of the friction. Putting together the two effects, one finds that the complete result (red curves) is very close to the case with no thermal masses, with differences at most of order $\textrm{few}\,\%$.

\begin{figure}[t]
\centering
\includegraphics[width=.95\linewidth]{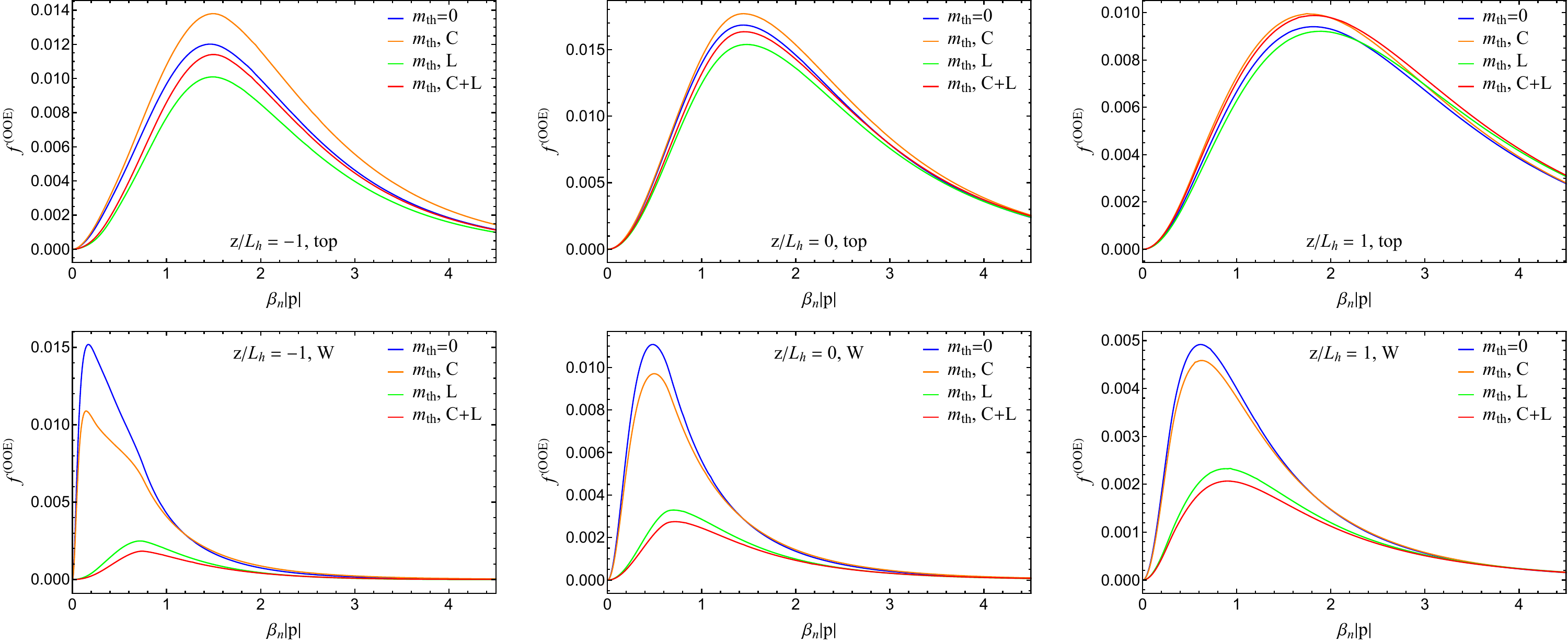}
    \caption{Friction integrands for BP$1$. The upper (lower) row shows the top-quark ($W$-boson) contribution. From left to right, the columns correspond to $z/L_h=-1$, $0$, and $1$.}
    \label{fig: friction integrand BP1}
\end{figure}

\begin{figure}
    \centering
    \includegraphics[width=.95\linewidth]{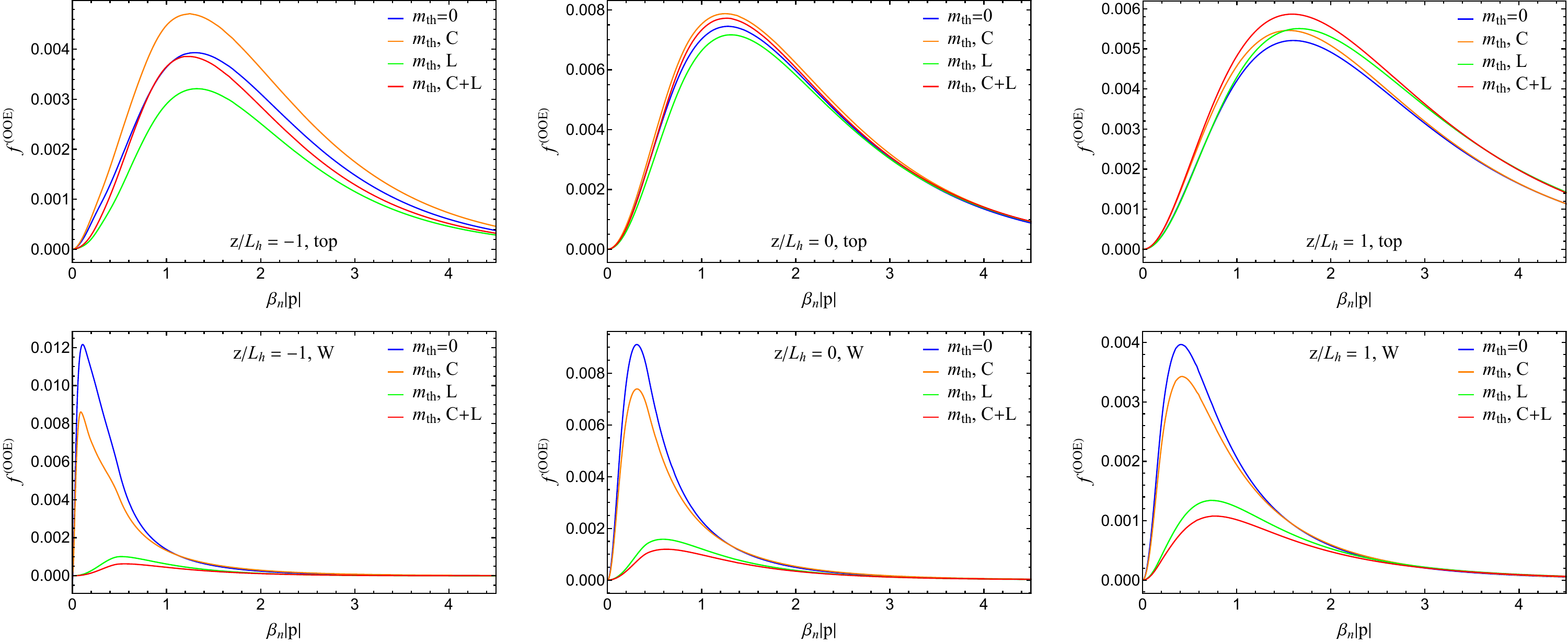}
    \caption{Same as Fig.~\ref{fig: friction integrand BP1}, but for BP$2$.}
    \label{fig: friction integrand BP2}
\end{figure}

In Fig.~\ref{fig: friction integrand BP1} and \ref{fig: friction integrand BP2} we show the friction integrand $f_t^{\scriptscriptstyle\rm (OOE)}$ of the top quark and $f_{_W}^{\scriptscriptstyle\rm (OOE)}$ of the $W$ bosons on the solutions where the latter are also included. The definition of $f_{_W}^{\scriptscriptstyle\rm (OOE)}$ is analogous to the one for $f_t^{\scriptscriptstyle\rm (OOE)}$ given in \eqref{eq: f_t OOE}.


The most striking impact of thermal masses can be seen in the $W$ boson contributions. With no thermal masses the friction is dominated by soft gauge bosons and receives a large enhancement from the infrared behaviour of the Bose-Einstein distribution. This is particularly evident outside the bubble, where the friction integrand peaks at $\beta_n |\vec p\,| \simeq 0.25$. Once thermal masses are included, this enhancement is removed and the dominant contribution shifts towards momenta $\beta_n |\vec p\,| \sim 1$, where the quasiparticle picture underlying the Boltzmann equation is expected to provide a reliable description. The friction is therefore controlled by momentum modes for which the kinetic treatment is on much firmer theoretical grounds. At the same time, the overall magnitude of the $W$-boson contribution is significantly reduced, rendering it subleading with respect to the top-quark contribution. The determination of the friction is thus not only quantitatively modified, but also considerably less sensitive to the poorly controlled infrared sector of the plasma.

As expected, the dominant origin of reduction is the suppression of the source term once thermal masses are included in the Liouville operator, whereas the modification of the collision integrals has a much milder impact. It is nevertheless interesting to notice that
the inclusion of thermal masses in the collision integrals induces a mild suppression of $f_{_W}^{\scriptscriptstyle\rm (OOE)}$ at low momenta, which is due to the small enhancement of the $W$ scattering kernel visible in Fig.~\ref{fig: Gl top}. Compared to the dramatic effect associated with the Liouville operator, this modification
remains quantitatively small and does not alter the overall picture.

We also notice that the inclusion of the $W$-boson contributions modifies the impact of thermal masses on the top-quark friction. In particular, inside the bubble the complete treatment with thermal masses leads to a mild enhancement of the top contribution. This is not a direct effect of the thermal masses on the top-quark Boltzmann equation, but rather an indirect consequence of the modifications in the domain-wall speed and in the temperature and fluid-velocity profiles. We will return to this point when discussing the plasma profiles below.

\begin{figure}[t]
\centering
\includegraphics[width=.95\linewidth]{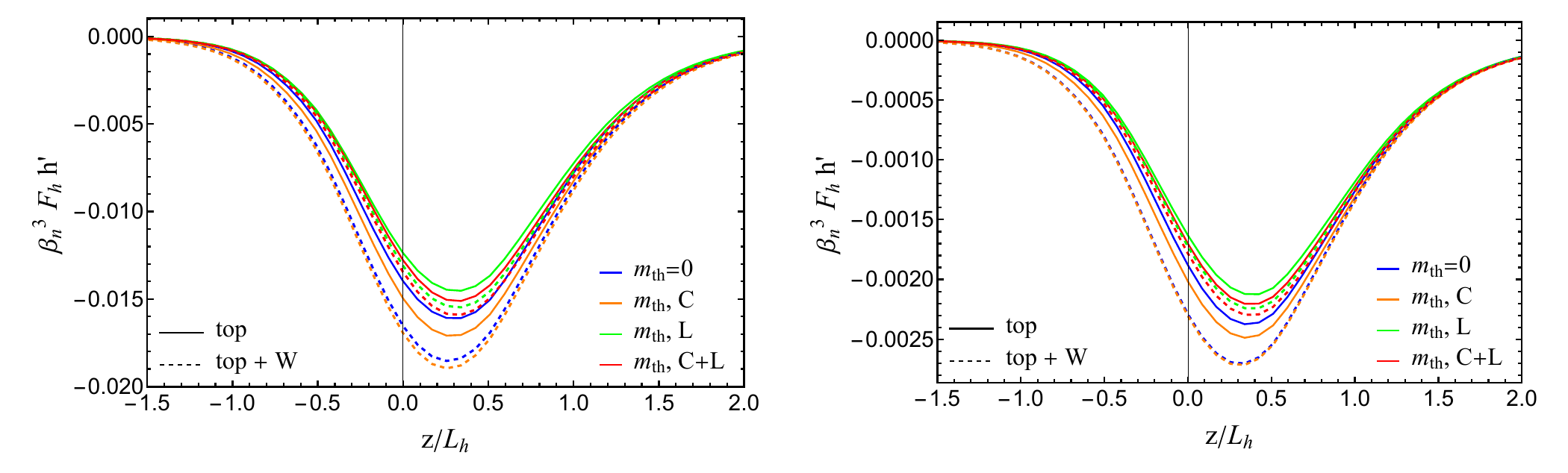}
\caption{Profiles of the out-of-equilibrium friction $F_h^{\scriptscriptstyle \rm OOE}(z) h'(z)$ for the two benchmark points BP1 (left panel) and BP2 (right panel), for the different treatments of thermal masses discussed in the text.}
    \label{fig: friction benchmarks}
\end{figure}

The impact of thermal masses can also be appreciated by analysing the out-of-equilibrium friction as a function of the coordinate $z$, as sown in Fig.~\ref{fig: friction benchmarks}. Obviously, the dominant contributions come from the region close to the domain wall and show a peak slightly inside the bubble, at $z/L_h \simeq 0.3$. Thermal masses affect the friction smoothly throughout the wall, inducing only a moderate distortion of the shape, more pronounced in front of the wall. Considerations similar to those above on the size of $W$-boson contributions with respect to the top quark ones in the various set-ups can be drawn.

\begin{figure}[t]
\centering
\includegraphics[width=.95\linewidth]{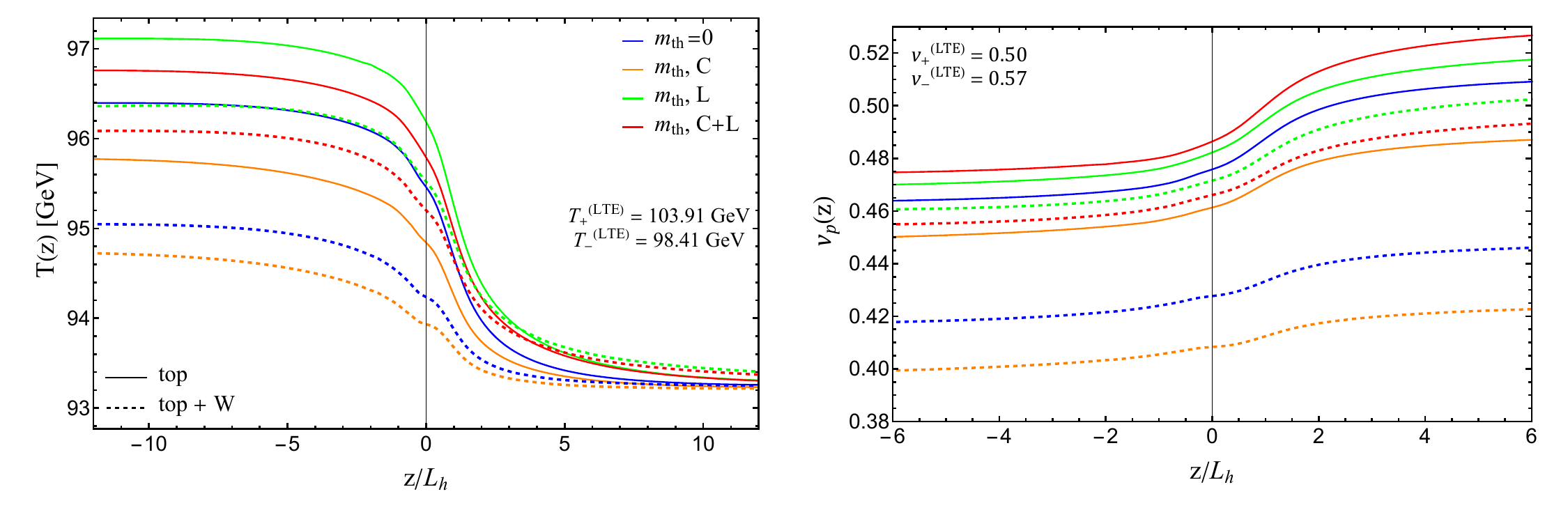}
\includegraphics[width=.95\linewidth]{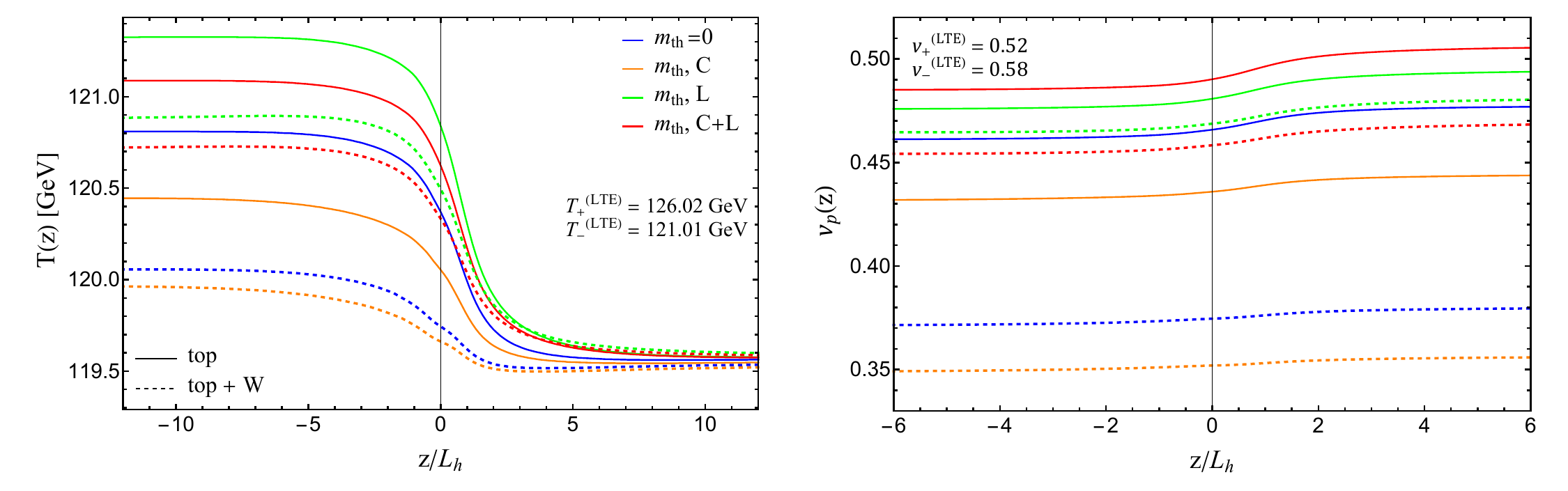}
\caption{Temperature profiles (left panels) and plasma-velocity profiles (right panels) for the two benchmark points. BP$1$ is shown in the upper row and BP$2$ in the lower row.}
\label{fig: profiles benchmarks}
\end{figure}

We conclude by briefly discussing the temperature and plasma-velocity profiles shown in Fig.~\ref{fig: profiles benchmarks}. Compared with the LTE approximation (in the plots we only give the asymptotic values $T_{\pm}^{\textsc{lte}}$ and $v_{\pm}^{\textsc{lte}}$), the inclusion of out-of-equilibrium friction reduces the gradients of both $T(z)$ and $v_p(z)$. This behaviour follows directly from the fact that part of the pressure balance across the wall is now provided by the out-of-equilibrium force, reducing the amount of hydrodynamic friction that must be generated through the temperature gradient\footnote{As in LTE $\gamma(z)T(z)={\rm const}$, the temperature gradient induces a velocity gradient, and the two are related.}.

The impact of thermal masses on the plasma profiles mirrors the behaviour already observed for the wall velocity and the friction integrands. In particular, the inclusion of thermal masses strongly suppresses the out-of-equilibrium contribution from $W$ bosons, reducing their influence on the hydrodynamic solution. As a result, the differences between the solutions obtained with and without $W$ bosons become considerably smaller once thermal masses are included. Overall, the modifications of the temperature and velocity profiles remain moderate and are consistent with the relatively small changes observed in the wall velocity.

\subsection{Impact across the parameter space}\label{sec:scan}

\begin{figure}[t]
\centering
\includegraphics[width=\linewidth]{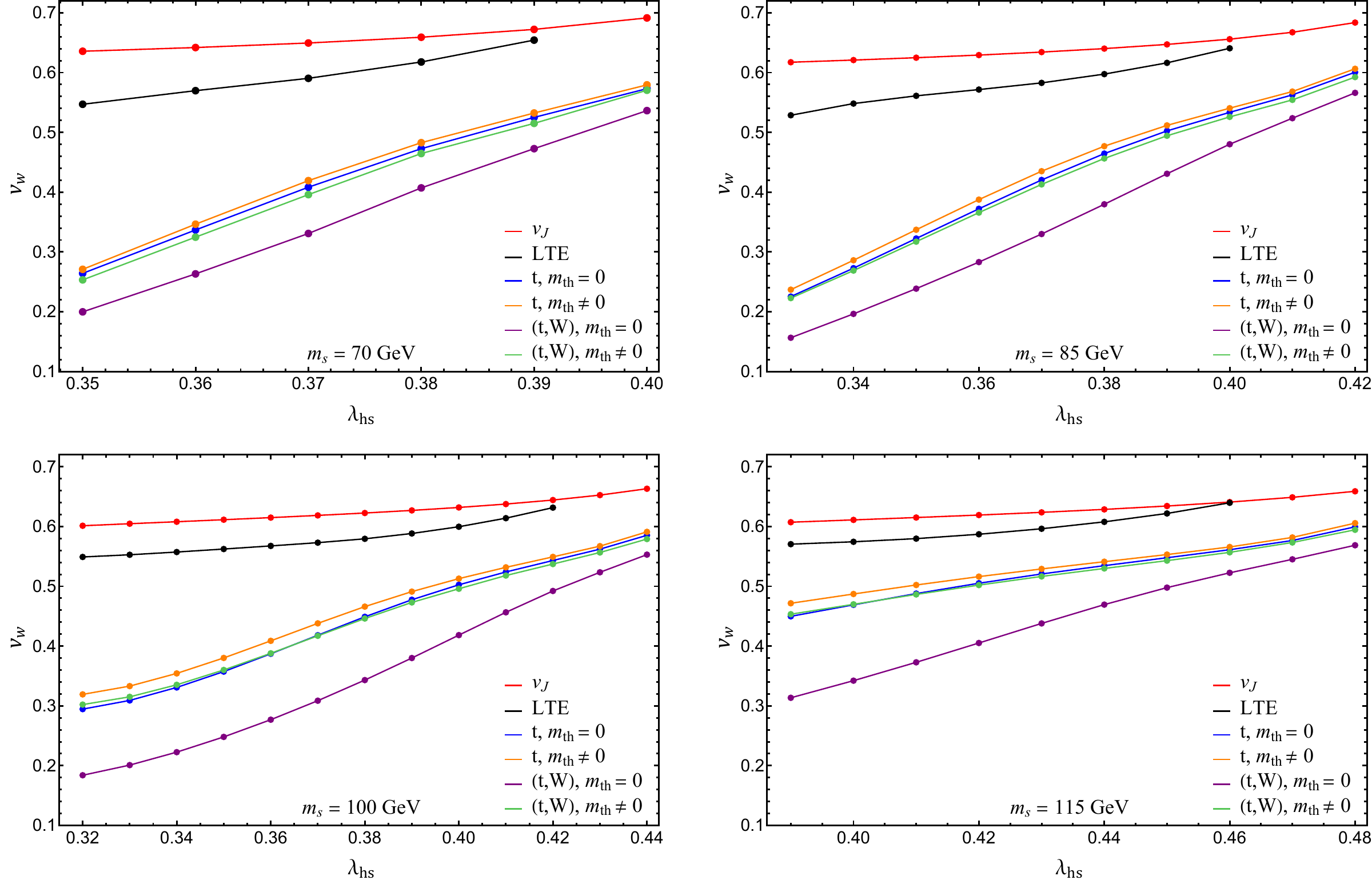}
    \caption{Wall velocity $v_w$ as a function of the portal coupling $\lambda_{hs}$ for different values of the singlet mass $m_s$ and fixed self-coupling $\lambda_s=1$. For the cases with $m_{\rm th}\ne 0$, thermal masses are included in both $\mathcal C$ and $\mathcal L$. The quantity $v_{J}$ denotes the Jouguet velocity. 
    }
    \label{fig: vw lines}
\end{figure}

As a final step of our analysis, we investigate how thermal masses affect the wall velocity throughout the parameter space of the SSM. To obtain a representative sampling of the parameter space, we consider four values of the singlet mass $m_s$, fix $\lambda_s=1$ and vary $\lambda_{hs}$ within the region with a two-step phase transition~\cite{Branchina:2025jou,Branchina:2025adj}. The results found for the domain wall speed $v_w$ are shown in Fig.~\ref{fig: vw lines}.

The most important observation is that the qualitative picture found for the benchmark points persists throughout the parameter space explored. When we consider the friction generated by the top quark alone, the wall velocities obtained in the massless approximation and in the treatment including thermal masses differ only mildly. This confirms that, over the region considered, the effects associated with thermal masses largely compensate in the top sector.

A second robust feature is that the inclusion of thermal masses significantly reduces the contribution from $W$ bosons. The corresponding wall velocities are always close to those obtained in the top-only approximation. This indicates that the $W$-boson contribution to the friction becomes subleading once the infrared enhancement present in the massless approximation is regulated. The scan shows that this behaviour is not restricted to isolated benchmark points but persists throughout the parameter region considered.

\begin{table}[t]
\centering
{\small
\begin{tabular}{@{}c|c|c|c|c@{}}
$m_s$ [GeV] & $t,\,m_{\rm th}=0$ & $t,\,m_{\rm th}\ne0$ & $(t,W),\,m_{\rm th}=0$ & $(t,W),\,m_{\rm th}\ne0$  \\
\hline
\rule{0pt}{1.1em} 70 & 51.7\% - 19.8\% & 50.5\% - 18.6\% & 63.5\% - 27.7\% &53.7\% - 21.3\%\\ 
\rule{0pt}{1.1em} 85 & 57.4\% - 16.7\% & 55.2\% - 15.6\% & 70.4\% - 25.0\% &57.9\% - 17.9\% \\ 
\rule{0pt}{1.em} 100 & 46.4\% - 14.0\% & 41.9\% - 13.0\% & 66.5\% - 22.1\% & 45.0\% - 14.9\% \\ 
\rule{0pt}{1.1em} 115 & 21.2\% - 12.3\% & 17.3\% - 11.6\% & 45.0\% - 18.3\% & 20.5\% - 13.0\%  \\ 
\end{tabular}
}
\caption{Deviation $\delta v_w$ of the wall velocity from its LTE value for the first and last parameter points admitting an LTE solution in each panel of Fig.~\ref{fig: vw lines}
($m_s = 70,\, 85,\, 100,\, 115$ GeV) and for the different treatments of thermal masses.}
\label{tab: deviations lines}
\end{table}

The figure also illustrates a feature already discussed in Refs.~\cite{Branchina:2025jou,Branchina:2025adj}. In LTE, stationary solutions exist only for wall velocities below the Jouguet velocity, $v_w \leq v_J$, which restricts the region of parameter space where solutions can be found. The inclusion of out-of-equilibrium friction enlarges the region of parameter space in which stationary deflagration solutions exist. As a result, some points shown in Fig.~\ref{fig: vw lines} admit a non-equilibrium solution even though no LTE solution exists.

Finally, Table~\ref{tab: deviations lines} provides a
measure of the impact of OOE corrections across the parameter space. For each value of $m_s$ and each treatment of thermal masses, we report the relative deviation of the wall velocity from its LTE value, $\delta v_w = 1-{v_w}/{v_w^{\textsc{lte}}}$. In each column, the values shown correspond to the first and last parameter points for which an LTE solution exists.

\section{Conclusions}\label{sec:conclusions}

We investigated the impact of thermal masses on the out-of-equilibrium friction acting on expanding bubble walls during a first-order electroweak phase transition. As a concrete case study, we focused on the singlet-extended Standard Model, which admits a two-step electroweak transition.

Thermal masses modify the Boltzmann equation describing the plasma perturbations by altering the equilibrium distribution functions and interaction rates. On the one hand, they reduce the source term $\mathcal{S}$ generated by the Liouville operator acting on the local-equilibrium distributions. On the other hand, they suppress the collision rates, slowing thermalisation and making out-of-equilibrium perturbations more persistent.

We found that in the top sector the two effects largely compensate, leading to only percent-level changes in the wall velocity with respect to the approximation neglecting thermal masses. The effects are clearly visible in the two benchmarks analysed in Sec.~\ref{sec:benchmarks}, where thermal masses induce only mild modifications of the friction profiles while preserving their overall shape and features.

The situation is qualitatively different for the $W$ bosons. In the massless approximation their contribution is dominated by soft momentum modes enhanced by the Bose-Einstein distribution, as visible in the lower panels of Figs.~\ref{fig: friction integrand BP1} and \ref{fig: friction integrand BP2}. For gauge bosons, the dominant effect of thermal masses is the suppression of the source term $\mathcal{S}$, which strongly reduces the contribution from soft modes. As a consequence, the infrared enhancement of the friction is regulated, the low-momentum peak in the friction integrand disappears, and the dominant contribution shifts towards momenta of order $T$. This results in a substantial suppression of the overall $W$-boson friction.

Once thermal masses are included, the gauge-boson contribution becomes subleading with respect to the top-quark one. The resulting wall velocities and the other parameters characterising the bubble-wall dynamics are therefore close to those obtained in the top-only approximation.

As shown in Sec.~\ref{sec:scan}, this qualitative picture remains valid across the parameter space of the SSM. Remarkably, throughout the parameter region explored, the approximation retaining only the top contribution to the friction and neglecting thermal masses remains very close to the full result, with deviations of only a few percent (see Fig.~\ref{fig: vw lines}).

Beyond their quantitative impact on the wall velocity, our results show that the inclusion of thermal masses substantially reduces the sensitivity of friction calculations to the infrared sector of the plasma, where the assumptions underlying the Boltzmann description are least reliable. This provides a firmer theoretical basis for determining bubble-wall velocities in first-order phase transitions. In this sense, thermal masses not only modify the predicted friction, but also improve the theoretical control over its determination.

\section*{Acknowledgments}
CB acknowledges support by the Deutsche Forschungsgemein- schaft (DFG, German Research
Foundation) under grant 396021762 - TRR 257.
The work has also been funded by the European Union – Next Generation EU
through the research grant number P2022Z4P4B “SOPHYA - Sustainable Optimised PHYsics
Algorithms: fundamental physics to build an advanced society” under the program PRIN 2022
PNRR of the Italian Ministero dell’Universit\`{a} e Ricerca (MUR) and by the research grant number 20227S3M3B “Bubble Dynamics in Cosmological Phase Transitions” under the program PRIN 2022 of the Italian Ministero dell’Universit\`{a} e Ricerca (MUR). MS is supported by a PhD studentship jointly funded by STFC and the University of Sussex. SDC would like to thank the Galileo Galilei Institute for Theoretical Physics (GGI) for the hospitality. AN thanks the Galileo Galilei Institute for Theoretical Physics (GGI) for hospitality within the Simons Visiting Program.

\end{document}